\documentclass[12pt,journal,onecolumn,draftcls]{IEEEtran}

\usepackage{epsfig}
\usepackage{times}
\usepackage{float}
\usepackage{afterpage}
\usepackage{amstext}
\usepackage{amssymb,bm}
\usepackage{latexsym}
\usepackage{color}
\usepackage{amsmath}
\usepackage{theorem}
\usepackage{stfloats}
\usepackage{pstricks}
\usepackage{subfigure}
\usepackage{enumerate}

\newcommand{\Rp}{R_\mathsf{p}}
\newcommand{\Rc}{R_\mathsf{c}}

\newcommand{\Xc}{{X_{\mathsf{c}}}}
\newcommand{\Xp}{{X_{\mathsf{p}}}}
\newcommand{\Tc}{{T_{\mathsf{c}}}}
\newcommand{\Tp}{{T_{\mathsf{p}}}}
\newcommand{\Yf}{{T_{\mathsf{f}}}}
\newcommand{\Yc}{{Y_{\mathsf{c}}}}
\newcommand{\Yp}{{Y_{\mathsf{p}}}}
\newcommand{\Zf}{{Z_{\mathsf{f}}}}
\newcommand{\Zp}{{Z_{\mathsf{p}}}}
\newcommand{\Zc}{{Z_{\mathsf{c}}}}

\newcommand{\Wp}{{W_{\mathsf{p}}}}
\newcommand{\Wc}{{W_{\mathsf{c}}}}
\newcommand{\snr}{\mathsf{S}}
\newcommand{\inr}{\mathsf{I}}
\newcommand{\Sp}{{\snr}_{\mathsf{p}}}
\newcommand{\Sc}{{\snr}_{\mathsf{c}}}
\newcommand{\Ip}{{\inr}_{\mathsf{p}}}
\newcommand{\Ic}{{\inr}_{\mathsf{c}}}
\newcommand{\Cc}{\mathsf{C}}
\newcommand{\eac}{{\rm e}^{{\rm j}\theta_{\mathsf{c}}}}
\newcommand{\eap}{{\rm e}^{{\rm j}\theta_{\mathsf{p}}}}

\newcommand{\gdof}{\mathsf{d}}

\newtheorem{theorem}{Theorem}
\newtheorem{lemma}{Lemma}
\newtheorem{remark}{Remark}
\newtheorem{defi}{Definition}

\allowdisplaybreaks

\title{The Two-user Causal Cognitive Interference Channel: 
Novel Outer Bounds and Constant Gap Result for the Symmetric Gaussian Noise Channel in Weak Interference}

\author{Martina~Cardone, Daniela~Tuninetti and Raymond~Knopp
\thanks{
M.~Cardone and R.~Knopp are with the Mobile Communications Department at Eurecom, Biot, 06410, France (e-mail: cardone@eurecom.fr; knopp@eurecom.fr). 
Eurecom's research is partially supported by its industrial partners: BMW Group Research \& Technology, IABG, Monaco Telecom, Orange, SAP, SFR, ST Microelectronics, Swisscom and Symantec. The research at Eurecom leading to these results has received funding from the EU Celtic+ Framework Program Project SHARING and from a 2014 Qualcomm Innovation Fellowship.
D. Tuninetti is with the Electrical and Computer Engineering Department of the University of Illinois at Chicago, Chicago, IL 60607 USA (e-mail: danielat@uic.edu).
The work of D.~Tuninetti was partially funded by NSF under award number 1218635;
the contents of this article are solely the responsibility of the author and
do not necessarily represent the official views of the NSF.

The results in this paper have been presented in part to the 6th International Symposium on Communications, Control and Signal Processing (ISCCSP) \cite{OurISCCSP2014} and to the 2014 IEEE International Symposium on Information Theory (ISIT) \cite{OurISIT2014}.
}
}

\begin{document}
\maketitle

\begin{abstract}
This paper studies the two-user Causal Cognitive Interference Channel (CCIC),
where two transmitters aim to communicate independent messages to two different receivers via a common channel. One source, referred to as the {\it cognitive}, is {\it capable} of overhearing the other source, referred to as the {\it primary}, through a noisy in-band link and thus can assist in sending the primary's data.
Two novel outer bounds of the type $2\Rp+\Rc$ and $\Rp+2\Rc$ are derived for the class of injective semi-deterministic CCICs where the noises at the different source-destination pairs are independent.
An achievable rate region is derived based on Gelfand-Pinsker binning, superposition coding  and simultaneous decoding at the receivers.

The lower and outer bounds are then specialized to the practically relevant Gaussian noise case. 
The authors of this paper recently characterized to within a constant gap the capacity of the symmetric Gaussian CCIC in 
(a) the strong interference regime, and 
(b) for a subset of the weak interference regime when the cooperation link is larger than a given threshold. 
This work characterizes to within a constant gap the capacity for the symmetric Gaussian CCIC in the regime that was still open.
In particular, it is shown that the novel outer bounds are necessary to characterize the capacity to within a constant gap when the cooperation link is weaker than the direct links, 
that is, in this regime unilateral cooperation 
leaves some system resources underutilized.
\end{abstract}


\section{Introduction}
\label{sec:intro}
This work considers the two-user Causal Cognitive Interference Channel (CCIC), a wireless network where one {\it primary} source PTx and one {\it cognitive / capable} source CTx aim to reliably communicate with two different receivers, namely the PRx and the CRx, via a common channel. Differently from the classical non-cooperative IC, in the CCIC the CTx operates in full-duplex mode and is able to overhear the PTx through a noisy in-band link; the CTx can thus exploit this side information to boost the performance of the two (primary and cognitive) systems. 

The major feature of the CCIC is the concept of {\it causal cognition} / {\it source cooperation}, which represents both an interference management tool and a practical model for the cognitive radio technology.
Actually, unilateral source cooperation offers a way to smartly cope with interference. In today's wireless systems, the general approach to deal with interference is either to avoid it, by trying to orthogonalize (in time / frequency / space) users' transmission, or to treat it as noise. However, these approaches may severely limit the system capacity since perfect user orthogonalization is not possible in practice. In contrast, in the CCIC the CTx, which can causally learn the primary's data through a noisy link, may protect both its own (by precoding against some known interference) and the primary's (by allocating some of its transmission resources to assist the PTx to convey data to the PRx) information from interference. Thus, the transmission techniques designed for the CCIC aim to leverage the structure of the interference, instead of just simply disregarding it and treating it as noise. 
The CCIC also represents a more practically relevant model for the cognitive {\it overlay} paradigm, compared to the case where the CTx is assumed to a priori (before the transmission begins) know the message of the PTx \cite{Devroye}, which may be granted only in limited scenarios. In contrast, in the CCIC the CTx causally learns the PTx's data through a noisy link. Thus, the transmission techniques designed for the CCIC account for the time the CTx needs for decoding and for the (possible) further rate losses that may incur in decoding the PTx's message through a noisy link of limited capacity.

\subsection{Related past work}
The CCIC studied in this work models an IC with unilateral source cooperation and represents a practical scenario for cognitive radios, where one source has superior capabilities with respect to the other source. Moreover, closely related to the IC with unilateral source cooperation is the IC with perfect output feedback model, where the received signal is fed back through a perfect channel from one receiver to the corresponding transmitter. Lately, these scenarios have received significant attention, as summarized next.

\subsubsection{IC with source cooperation} 
The CCIC is a particular case of the IC with bilateral source cooperation, where one of the cooperation link (in our case the one from the CTx to the PTx) is absent. For the IC with bilateral source cooperation several outer bounds on the capacity region were derived in \cite{HostMadsenIT06,TuninettiITA10,PVIT11,TndonUlukusIT11} and a number of transmission strategies were designed in \cite{YANG-TUNINETTI}. 

In \cite{HostMadsenIT06}, the author firstly derived inner and outer bounds on the capacity for the IC with bilateral source cooperation and for the IC with bilateral destination cooperation. The outer bound region was obtained from the max-flow-min-cut theorem, by further strengthening the sum-rate bound $\Rp+\Rc$ (where $\Rp$, respectively $\Rc$, is the transmission rate for the PRx, respectively the CRx) as proposed in \cite{CaireShamai}, while the lower bound region was derived by designing a scheme based on Gelfand-Pinsker's binning \cite{gelfandpinsker} (i.e., dirty-paper-coding in Gaussian noise \cite{costaDPC}) and superposition encoding, decode-and-forward relaying and simultaneous decoding.

In \cite{TuninettiITA10}, the author derived a novel general outer bound for the IC with bilateral source cooperation that applies to any channel, i.e., the noises can be arbitrarily correlated. The outer bound rate region in \cite[Theorem II.1]{TuninettiITA10} has constraints on the single rates $\Rp$ and $\Rc$ and on the sum-rate $\Rp+\Rc$. In particular, the constraints on the sum-rate were obtained by extending the idea of Kramer in \cite[Theorem 1]{Kramer}, beyond the Gaussian noise case. Moreover, the outer bounds in \cite[Theorem II.1]{TuninettiITA10} were recently shown to hold for the case of bilateral destination cooperation as well \cite{TuninettiITW12}. When evaluated for the symmetric Gaussian noise case, the derived region is achievable to within $2$ bits in the strong cooperation regime \cite{YangHighCoop}.

In \cite{PVIT11}, the authors derived a novel sum-rate outer bound for a class of Injective Semi-Deterministic (ISD) ICs with bilateral source cooperation and with independent noises at all terminals; when evaluated for the Gaussian noise case with symmetric cooperation links, the derived region is achievable to within $19$ bits. 

In \cite{TndonUlukusIT11}, the authors derived a novel outer bound on the capacity region of the Gaussian IC with bilateral source cooperation that are based on the idea of `dependence balance' proposed in \cite{kekstrawillems}. In \cite{TndonUlukusIT11} it was proved that the novel bound region may be tighter than the cut-set bound. 

In \cite{YANG-TUNINETTI} two transmission strategies for the IC with bilateral source cooperation were designed. The two schemes employ partial-decode-and-forward relaying, rate splitting and simultaneous decoding. While the first strategy (see \cite[Section IV]{YANG-TUNINETTI}) uses only superposition coding, the second scheme (see \cite[Section V]{YANG-TUNINETTI}) also employs Gelfand-Pinsker's binning in the encoding phase.

The CCIC was specifically studied in \cite{MirmohseniIT2012}, where novel outer and inner bounds were derived. Although, to the best of our knowledge, the work in \cite{MirmohseniIT2012} gives the best known bounds for this channel,  
their evaluation 
is not straightforward. For example,
the outer bound involves several auxiliary random variables that are jointly distributed with the inputs and for which no cardinality bounds on the corresponding alphabets are known; the inner bound is characterized by many constraints and auxiliary random variables whose optimization is not immediate.

In \cite{ourITjournal}, the capacity of the Gaussian CCIC (GCCIC) was characterized to within a constant gap for a set of channel parameters that, roughly speaking, excludes the case of weak interference at both receivers. For the symmetric case (i.e., the two direct links and the two interfering links are of the same strength) a constant gap result of $1$ bit was proved in strong interference and in weak interference when the cooperation link is `sufficiently strong'.
Moreover, in \cite{ourITjournal}, the capacity of the GCCIC was characterized to within $2$ bits for two special cases: 
the Z-channel, where the CRx does not experience interference from the PTx, and 
the S-channel, where the PRx does not experience interference from the CTx. 
These constant gap results were obtained by using known outer bounds on the single rates $\Rp$ and $\Rc$ and on the sum-rate $\Rp + \Rc$.

In \cite{ourITjournal} we pointed out that in weak interference the capacity region of the GCCIC should also have outer bounds of the type $2 \Rp + \Rc$ and $\Rp + 2\Rc$, similarly to the classical non-cooperative IC \cite{etw}. To the best of our knowledge, these bounds are not available in the existing literature of cooperative ICs and their derivation represents the main contribution of this work. These novel bounds are proved to be active for the symmetric GCCIC in weak interference and when the cooperation link is weaker than the direct link, thus closing a problem left open in \cite{ourITjournal}.

\subsubsection{IC with output feedback}
In \cite{suhtse:ICwithfeedback}, the authors studied the Gaussian IC where each source has a perfect output feedback from the intended destination and characterized the capacity to within $2$ bits.
In \cite[Theorems 2-3]{suhtse:ICwithfeedback}, the capacity region has constraints on the single rates $\Rp$ and $\Rc$ and on the sum-rate $\Rp+\Rc$, but not bounds of the type $2\Rp + \Rc$ and $\Rp + 2\Rc$, which appear in the capacity region of the classical Gaussian IC in weak interference \cite{etw}. The authors interpreted the bounds on $2\Rp + \Rc$ and $\Rp + 2\Rc$ in the capacity region of the classical IC as a measure of the amount of `resource holes', or system under utilizations, due to the distributed nature of the non-cooperative IC. The authors thus concluded that output feedback eliminates these `resource holes', i.e., the system resources are fully utilized.

In \cite{SahaiIT2013}, the symmetric Gaussian IC with all $9$ possible output feedback configurations was analyzed. The authors proved that the bounds derived in \cite{suhtse:ICwithfeedback} suffice to approximately characterize the capacity of all the $9$ configurations except for the case where only one source receives feedback from the corresponding destination, i.e., the `single direct-link feedback model', or the feedback-model-(1000). For the feedback-model-(1000), it was shown that an outer bound of the type $2\Rp + \Rc$ is needed to capture the fact that the second source (whose transmission rate is $\Rc$) does not receive feedback. In the language of \cite{suhtse:ICwithfeedback} we thus have that a `single direct-link feedback' does not suffice to cover all the `resource holes' whose presence is captured by the bound on $2\Rp + \Rc$. 
Such a bound was shown to be active for the feedback-model-(1000) in the Gaussian noise case.

In \cite{TandonPoor}, the authors characterized the capacity of the two-user symmetric linear deterministic approximation of the Gaussian IC with bilateral noisy feedback, i.e., where only some of the received bits are fed back to the corresponding transmitter.  
In \cite{TandonPoorJournalGaussian}, the same authors evaluated the bounds for the symmetric Gaussian noise channel and proved that they are at most $11.7$ bits far from one another, universally over all channel parameters.
The capacity characterization was accomplished by proving novel outer bounds on $2\Rp + \Rc$ and $\Rp + 2\Rc$
that rely on carefully chosen side information random variables tailored to the symmetric Gaussian setting and whose generalization to non-symmetric or non-Gaussian scenarios does not appear straightforward.

In this work we first provide a general framework to derive outer bounds of the type $2\Rp + \Rc$ and $\Rp + 2\Rc$ for the ISD CCIC,  which recover as special cases those derived in \cite{SahaiIT2013,TandonPoor}.
We then show that these novel outer bounds are active for the symmetric GCCIC in weak interference when the cooperation link is weaker than the direct link, i.e., in this regime unilateral cooperation does not enable sufficient coordination among the sources for full utilization of the channel resources.

\subsubsection{Non-causal CIC}
In the original cognitive radio {\it overlay} paradigm, first introduced in \cite{Devroye}, the superior capabilities of the CTx were modeled by assuming that 
the CTx a priori knows the PTx's message.
The largest known achievable rate region for the general memoryless non-causal CIC 
is \cite[Theorem 7]{riniJ2}, which in \cite{riniJ1} was evaluated for the Gaussian noise case and shown to be
at most $1$ bit apart (see \cite[Theorem VI.1]{riniJ1}) from an outer bound region, which is characterized by constraints on the single rates $\Rp$ and $\Rc$ and on the sum-rate $\Rp+\Rc$. In other words, the capacity region of the non-causal CIC does not have bounds on $2\Rp+\Rc$ and $\Rp+2\Rc$, i.e., the assumption of full a priori knowledge of the PTx's message at the CTx allows to fully exploit the available system resources. 

In \cite{ourITjournal}, we removed the ideal non-causal message knowledge assumption by considering the CCIC and we identified the set of parameters where unilateral cooperation attains the ultimate performance limits (i.e., generalized Degrees of Freedom (gDoF)) predicted by the non-causal model. 
In this work we show that the capacity of the CCIC differs in general from that of the ideal non-causal CIC since its capacity region has bounds of the type $2\Rp+\Rc$ and $\Rp+2\Rc$, i.e., in general removing the ideal full a priori message knowledge at the CTx leaves some `resource holes' in the system.

\subsection{Contributions}
Our main contributions can be summarized as follows:
\begin{enumerate}
\item We develop a general framework to derive outer bounds of the type $2\Rp + \Rc$ and $\Rp + 2\Rc$ on the capacity of the general ISD CCIC when the noises at the different source-destination pairs are independent; this framework includes for example feedback from the intended destination. As a special case, we recover and strengthen the bounds derived in \cite{PVIT11,SahaiIT2013,TandonPoor}. The key technical ingredient is the proof of two novel Markov chains.
\item We design a transmission strategy for the general memoryless CCIC and we derive its achievable rate region. The proposed scheme uses superposition and binning encoding, partial-decode-and-forward relaying and simultaneous decoding at the receivers. Since the CCIC shares common features with the classical non-cooperative IC \cite{HK}, both {\it common} (decoded also at the non-intended receiver) and {\it private} (treated as noise at the non-intended receiver) messages are used. Moreover, we use both {\it cooperative} (first partially decoded, then re-encoded and finally forwarded by the CTx) and {\it non-cooperative} (sent without the help of the CTx) messages for the PTx, while the messages of the CTx are only {\it non-cooperative}.
\item We evaluate the outer bound and the achievable rate regions for the practically relevant Gaussian noise channel. We prove that for the symmetric case, i.e., when the two direct links and the two cross / interfering links are of the same strength, the achievable region is a constant (uniformly over all channel gains) number of bits apart from the outer bound region in the regimes that were left open in \cite{ourITjournal}, i.e., weak interference and weak cooperation. This result, jointly with \cite[Theorem 1]{ourITjournal}, fully characterizes the capacity of the symmetric GCCIC to within a constant gap. 
Moreover, this result sheds light on the regimes where unilateral cooperation is too weak and leaves some system resources underutilized.
\end{enumerate}

\subsection{Paper organization}
The rest of the paper is organized as follows. 
Section \ref{sec:channel model} describes the general memoryless CCIC, the ISD CCIC and the Gaussian CCIC, for which the concepts of capacity to within a constant gap and gDoF are defined. 
Section \ref{sec:outer} is dedicated to the outer bounds.
In Section \ref{sec:outerPast} we first summarize known outer bounds and then generalize and tighten the sum-rate bound in~\cite[bound $u_{1}$]{PVIT11}
by proving two novel Markov chains for the involved random variables.
Then, Section \ref{sec:outerNew} derives outer bounds of the type $2\Rp+\Rc$ and $\Rp+2\Rc$ for the ISD CCIC with independent noises at the different source-destination pairs. 
Section \ref{sec:gap sumcapacity Symmetric Channel} focuses on the Gaussian noise CCIC and shows that the two novel outer bounds suffice to prove a constant gap result for the symmetric case in weak interference when the cooperation link is not `strong enough', i.e., the regime which was left open in \cite{ourITjournal}. In particular, in Section \ref{subsect:OBGaussian} the outer bounds are evaluated for the GCCIC, in Section \ref{subsect:IBGaussian} a novel transmission strategy is designed and its achievable rate region is derived and finally in Section \ref{subsec:gapGaussian} the achievable rate region is shown to be a constant number of bits apart from the outer bound region.
Section \ref{sec:Conclusion} concludes the paper.
Some proofs may be found in Appendix. 

Although this paper considers only the symmetric Gaussian noise CCIC, we believe that the results can be extended to the general  non-symmetric case albeit with more tedious and involved computations, especially for the achievable region, than those reported here.

\subsection{Notation}
Throughout the paper we adopt the following notation convention. 
The subscript $\mathsf{c}$ (in sans serif font) is used for quantities related to the cognitive pair, while the subscript $\mathsf{p}$ (in sans serif font) for those related to the primary pair. The subscript $\mathsf{f}$ or $\mathsf{F}$ (in sans serif font) is used to refer to generalized feedback information received at the CTx.  
The subscript $c$ (in roman font) is used to denote both common and cooperative messages, the subscript $p$ (in roman font) to denote private messages and the subscript $n$ (in roman font) to denote non-cooperative messages.
With $[n_1:n_2]$ we denote the set of integers from $n_1$ to $n_2 \geq n_1$ and $[x]^+ := \max\{0,x\}$ for $x\in\mathbb{R}$;
$Y^{j}$ is a vector of length $j$ with components $(Y_1,\ldots,Y_j)$ and $\mathbb{E}\left [ \cdot \right ]$ indicates the expected value; $a^*$ denotes the complex conjugate of $a$ and $|a|$ is the absolute value of $a$; $\emptyset$ is the empty set. The notation eq$(n)$ is used to indicate the rightmost side of the equation number $n$.

\section{System Model}
\label{sec:channel model}

\subsection{The general memoryless channel}
\label{sec:GMCCIC}

\begin{figure}
\centering
\includegraphics[width=0.7\textwidth]{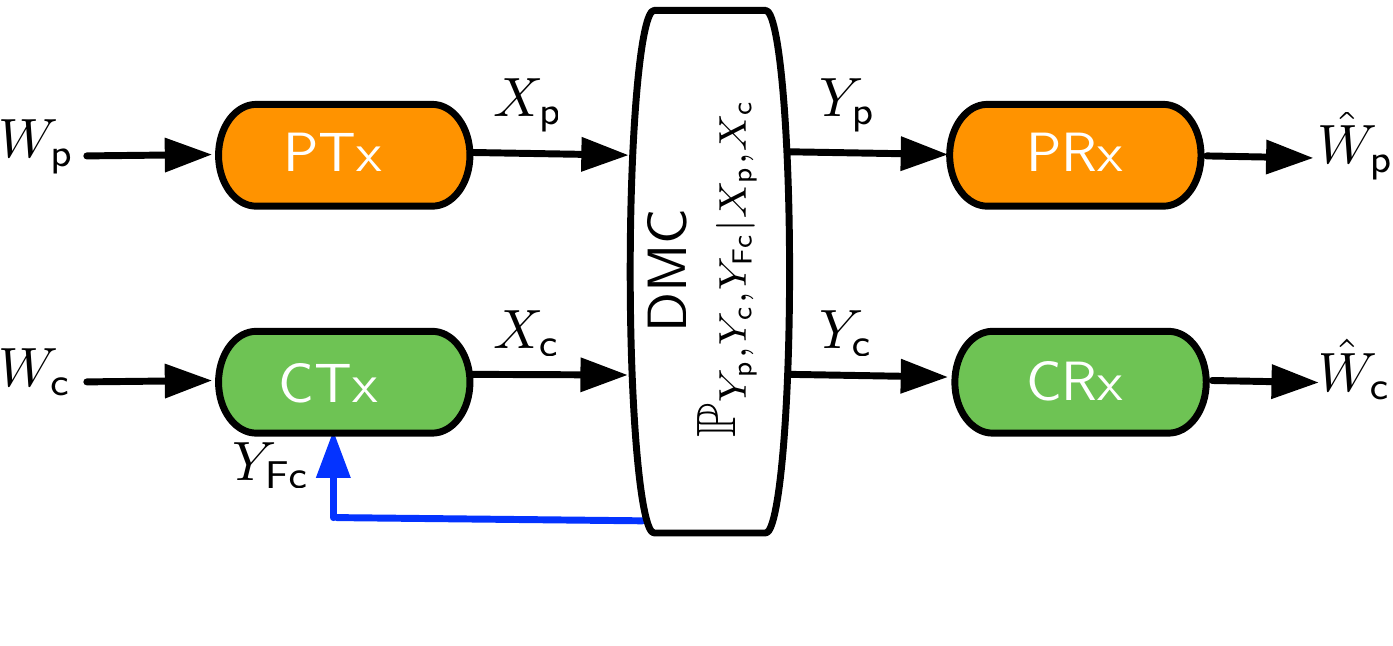}
\caption{The general memoryless CCIC.}
\label{fig:channelmodelDM}
\end{figure}

A {\em general memoryless CCIC}, shown in Fig.~\ref{fig:channelmodelDM}, consists of two input alphabets $\left (\mathcal{X}_{\mathsf{p}},\mathcal{X}_{\mathsf{c}} \right )$, three output alphabets $\left (\mathcal{Y}_{\mathsf{Fc}},\mathcal{Y}_{\mathsf{p}},\mathcal{Y}_{\mathsf{c}} \right )$ and a memoryless transition probability $P_{{Y}_{\mathsf{Fc}},\Yp,\Yc|\Xp,\Xc}$.
The PTx has a message $W_\mathsf{p}\in [1:2^{N \Rp}]$ for the PRx and the CTx has a message $W_\mathsf{c}\in [1:2^{N \Rc}]$ for the CRx, where $N\in\mathbb{N}$ denotes the codeword length and $\Rp\in\mathbb{R}_+$ and $\Rc\in\mathbb{R}_+$ the transmission rates for the PTx and the CTx, respectively, in bits per channel use (logarithms are in base 2).
The messages $W_\mathsf{p}$ and $W_\mathsf{c}$ are independent and uniformly distributed on their respective domains. 
At time $i, \ i\in [1:N],$ the PTx maps its message $W_\mathsf{p}$ into a channel input symbol $X_{\mathsf{p}i}(W_\mathsf{p})$ and the CTx maps its message $W_\mathsf{c}$ and its past channel observations into a channel input symbol $X_{\mathsf{c}i}(W_\mathsf{c},Y_{\mathsf{Fc}}^{i-1})$.
At time $N$, the PRx outputs an estimate of its intended message based on all its channel observations as $\widehat{W}_{\mathsf{p}}(\Yp^N)$, and similarly the CRx outputs $\widehat{W}_{\mathsf{c}}(\Yc^N)$.
The capacity region is the convex closure of all non-negative rate pairs $\left ( \Rp, \Rc \right )$ such that $\max_{u\in\{\mathsf{c},\mathsf{p}\}}\mathbb{P}[\widehat{W}_u \neq W_u] \to 0$ as $N\to+\infty$.

\subsection{The ISD channel}
\label{sec:ISDchannel}

\begin{figure}
\centering
\includegraphics[width=0.7\textwidth]{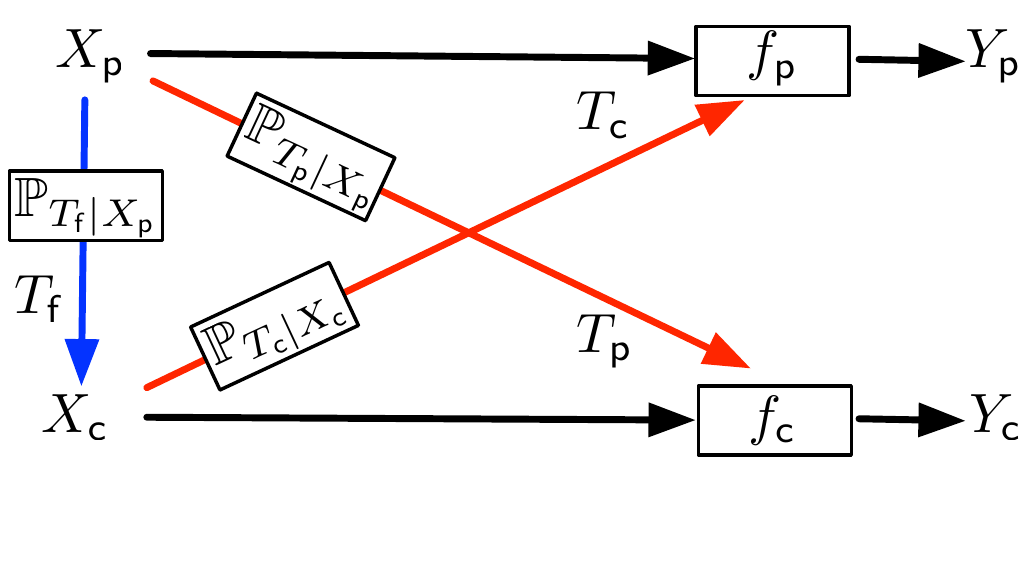}
\caption{The ISD CCIC.}
\label{fig:channelmodelISD}
\end{figure}

The {\em ISD} model, shown in Fig.~\ref{fig:channelmodelISD} and first introduced in \cite{TelatarTseISIT2007} for the classical IC, assumes that the input $\Xp$, respectively $\Xc$, before reaching the destinations, is passed through a memoryless channel to obtain $\Tp$, respectively $\Tc$. The channel outputs are therefore given by
\begin{subequations}
\begin{align}
   \Yp & = f_{\mathsf{p}} \left( \Xp,\Tc \right),
\\ \Yc & = f_{\mathsf{c}} \left( \Xc,\Tp \right),
\end{align}
where $f_u$, $u \in \{\mathsf{p},\mathsf{c}\}$, is a deterministic function that is invertible given $X_u$, or in other words, $\Tp$, respectively $\Tc$, is a deterministic function of $(\Yc,\Xc)$, respectively $(\Yp,\Xp)$.

In the CCIC, 
the generalized feedback signal at the CTx satisfies 
\begin{align}
{Y}_{\mathsf{Fc}} = f_{\mathsf{f}} \left( \Xc,\Yf \right),
\end{align}
\label{eq:injsemidetMod}
\end{subequations} 
for some deterministic function $f_{\mathsf{f}}$ that is invertible given $\Xc$, i.e., $\Yf$ is a deterministic function of $\left({Y}_{\mathsf{Fc}},\Xc \right)$, where $\Yf$ is obtained by passing $\Xp$ through a noisy channel \cite{PVIT11}.

In this work we assume that the noises seen by the different source-destination pairs are independent, that is,
\begin{align}
\mathbb{P}_{{Y}_{\mathsf{Fc}},\Yp,\Yc|\Xp,\Xc}
=
\mathbb{P}_{{\Yp|\Xp,\Xc}}
\mathbb{P}_{{Y}_{\mathsf{Fc}},\Yc|\Xp,\Xc}.
\label{eq:distr}
\end{align}
In other words, we assume that the noises at the PRx and at the CRx are independent, but we do not impose any constraint on the noises at the CTx and CRx. For example, the case of output feedback in \cite[model-(1000)]{SahaiIT2013} is obtained by setting $\Yf=\Tp$, which would not have been possible within the `all noises are independent' setting studied in  \cite{PVIT11}.

\subsection{The Gaussian noise channel}
\label{sec:ChGaussian}

\begin{figure}
\centering
\includegraphics[width=0.55\textwidth]{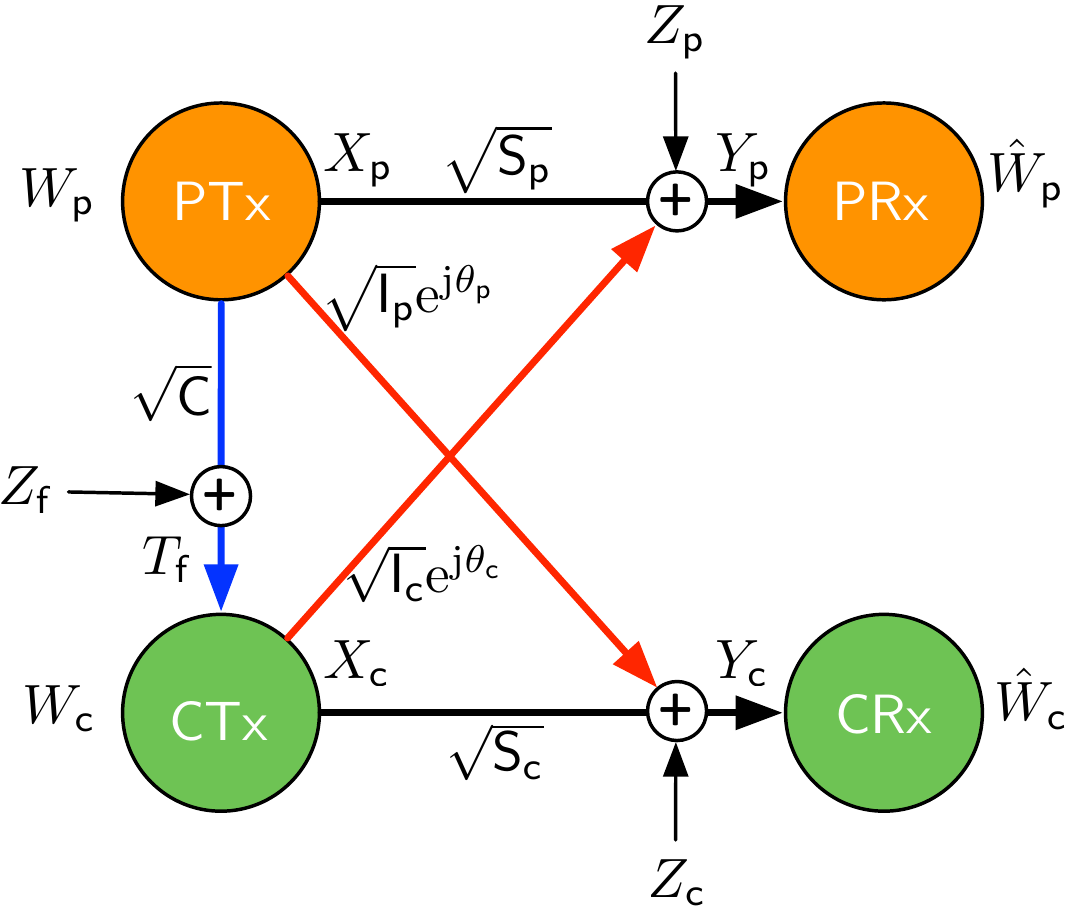}
\caption{The GCCIC.}
\label{fig:channelmodel}
\end{figure}

A single-antenna GCCIC, shown in Fig.~\ref{fig:channelmodel}, is a special case of the ISD model described in \eqref{eq:injsemidetMod} and it is defined by the input / output relationship
\begin{subequations}
\begin{align}
   \Tp &:= \sqrt{\Ip} \eap \Xp + \Zc, \label{eq:awgn tp}
\\ \Tc &:= \sqrt{\Ic} \eac \Xc + \Zp, \label{eq:awgn tc}
\\ \Yf & =  \sqrt{\Cc} \Xp + \Zf, \label{eq:awgn tf}
\\ \Yp & = \sqrt{\Sp}  \Xp + \Tc, \label{eq:awgn yp}
\\ \Yc & = \Tp + \sqrt{\Sc} \Xc, \label{eq:awgn yc}
\\ {Y}_{\mathsf{Fc}}& =  \Yf, \label{eq:awgn yf}
\end{align}
\label{eq:awgn full}
\end{subequations}
where $\Yf= {Y}_{\mathsf{Fc}}$ in~\eqref{eq:awgn yf} is without loss of generality since the CTx can remove the contribution of its transmit signal $\Xc$ from its received signal ${Y}_{\mathsf{Fc}}$.
The channel gains are assumed to be constant for the whole transmission duration, and hence known to all nodes. 
Without loss of generality, certain channel gains can be taken to be real-valued and non-negative since a node can compensate for the phase of one of its channel gains.
The channel inputs are subject to a unitary average power constraint, i.e., $\mathbb{E} \left [ |X_i|^2 \right ] \leq 1, i \in \{ \mathsf{p},\mathsf{c} \}$. This assumption is without loss of generality, since non-unitary power constraints can be incorporated into the channel gains. The noises are circularly symmetric Gaussian random variables with, without loss of generality, zero mean and unitary variance.
We assume that the noise $\Zp$ is independent of $(\Zc , \Zf )$, while $(\Zc,\Zf)$ can be arbitrarily correlated.

For the Gaussian noise case, it is customary to approximate the channel capacity as follows.

\begin{defi}
[Capacity region to within a constant gap]
The capacity region of the GCCIC is said to be known to within $\mathsf{GAP}$~bits if one can show an inner bound region $\mathcal{I}$ and an outer bound region $\mathcal{O}$ such that 
\[
(\Rp,\Rc)\in\mathcal{O} \Longrightarrow ([\Rp-\mathsf{GAP}]^+,[\Rc-\mathsf{GAP}]^+)\in\mathcal{I}.
\]
\end{defi}
For the two particular cases of $\Cc=0$ (i.e., non-cooperative IC) and of $\Cc\to+\infty$ (non-causal CIC), the capacity is known to within $1$~bit \cite{etw,riniJ1}. 

The approximate (i.e., to within a constant gap) characterization of the capacity region implies the exact knowledge of its gDoF region. 
The gDoF metric, first introduced in \cite{etw} for the non-cooperative IC, captures the high-SNR behavior of the capacity as a function of the relative strengths of the direct, cooperation and interfering links. The gDoF represents a more refined characterization of the capacity in the high-SNR regime compared to the classical DoF since it captures the fact that, in wireless networks, the channel gains can differ by several orders of magnitude. 
In this work, we consider the symmetric case parameterized as
\begin{subequations}
\begin{align}
     \Sp  = \Sc &:= {\snr}^1,        \ {\snr}\geq 0,   \ \text{direct links},
\\   \Ip  = \Ic &:= {\snr}^{\alpha}, \ \alpha\geq 0,   \ \text{interfering links},
\\         \Cc  &:= {\snr}^\beta,    \ \beta\geq 0,    \ \text{cooperation link},
\end{align}
\label{eq:paramet}
\end{subequations}
where $\alpha$ measures the strength of the interference links compared to the direct link, while $\beta$  the strength of the cooperation link compared to the direct link. Thus, the symmetric GCCIC is parameterized by the triplet $({\snr},\alpha,\beta)$, where ${\snr}$ is referred to as the (direct link) SNR, $\alpha$ as the interference exponent and $\beta$ as the cooperation exponent.\footnote{In principle the system performance also depends on the phases of the interfering links $(\theta_\mathsf{c},\theta_\mathsf{p})$. However, as far as gDoF and capacity to within a constant gap are concerned, the phases $(\theta_\mathsf{c},\theta_\mathsf{p})$ only matter if the IC channel matrix
${\small \begin{bmatrix}
\sqrt{\Sp} & \sqrt{\Ic} \eac \\
\sqrt{\Ip} \eap & \sqrt{\Sc} \\
\end{bmatrix}}$
is rank deficient \cite{suhtse:ICwithfeedback}, in which case one received signal is a noisier version of the other.
In this work, we assume that the phases are such that the IC channel matrix is full rank as in \cite{suhtse:ICwithfeedback}.}

\begin{defi}
[gDoF]
Given the parameterization in~\eqref{eq:paramet}, the gDoF is defined as
\begin{align}
{\gdof}(\alpha,\beta) 
&:= \lim_{{\snr}\to+\infty} \frac{\max\{\Rp+\Rc\}}{2\log(1+{\snr})},
\label{eq:gDoF definition}
\end{align}
where the maximization is intended over all possible achievable rate pairs $(\Rp,\Rc)$. 
\end{defi}

The gDoF of the classical IC ($\Cc=0$) is the ``W-curve'' first characterized in~\cite{etw} and given by ${\gdof}(\alpha,0)$.
The gDoF of the non-causal CIC ($\Cc\to\infty$) is the ``V-curve'', which can be evaluated from the capacity characterization to within 1 bit of~\cite{riniJ1}, and is given by ${\gdof}(\alpha,\infty)$. An interesting question this work answers is whether there are
 values of $\beta > 0$ such that ${\gdof}(\alpha,\beta)={\gdof}(\alpha,0)$ --- in which case unilateral causal cooperation is not helpful in terms of gDoF --- or values of $\beta < \infty$ such that ${\gdof}(\alpha,\beta)={\gdof}(\alpha,\infty)$ --- in which case unilateral causal cooperation is equivalent to non-causal message knowledge in terms of gDoF.

Following the naming convention of the non-cooperative IC \cite{etw}, we say that the symmetric GCCIC has strong interference if ${\snr} \leq {\inr}$, that is $1\leq \alpha$, and weak interference otherwise. Similarly, we say that the symmetric GCCIC has strong cooperation if ${\snr} \leq {\Cc}$, that is $1\leq \beta$, and weak cooperation otherwise.

\section{Outer bounds on the capacity region for the CCIC}
\label{sec:outer}
This section is dedicated to the study of outer bounds on the capacity region for the CCIC. 
First, in Section \ref{sec:outerPast}, some known outer bounds  are summarized. Moreover, the outer bound originally derived in \cite{PVIT11} for the ISD CCIC with independent noises at all terminals, is generalized to the case where only the noises at the different source-destination pairs are independent as in~\eqref{eq:distr}. 
Then, in Section \ref{sec:outerNew}, two novel outer bounds of the type $\Rp+2\Rc$ and $2\Rp+\Rc$ are derived for the ISD CCIC with independent noises at the different source-destination pairs as in~\eqref{eq:distr}.
As we shall see in Section~\ref{sec:gap sumcapacity Symmetric Channel}, these novel outer bounds allow to characterize the capacity to within a constant gap for the symmetric GCCIC in the regimes that were left open in \cite{ourITjournal}.

\subsection{Known outer bounds and some generalizations}
\label{sec:outerPast}
In the literature, several outer bounds are known for the IC with bilateral source cooperation~\cite{TuninettiITA10,PVIT11}, which we specialize here to the CCIC.
In particular, for a joint input distribution $\mathbb{P}_{\Xp,\Xc}$, we have:
\begin{enumerate}
\label{eq:knownOBgeneralmemo}
\item For the {\em general memoryless CCIC}, described in Section \ref{sec:GMCCIC}, the cut-set upper bound \cite{ElGamalKimBook} gives
\begin{subequations}
\label{eq:outknownbefore}
\begin{align}
   \Rp & \leq I \left( \Xp;\Yp,{Y}_{\mathsf{Fc}}|\Xc \right), \label{eq:cutset1 a}
\\ \Rp & \leq I \left( \Xp,\Xc; \Yp \right), \label{eq:cutset1 b}
\\ \Rc & \leq I \left( \Xc;\Yc|\Xp \right), \label{eq:cutset2}
\end{align}
and from~\cite{TuninettiITA10} we have
\begin{align}
   \Rp + \Rc & \leq I \left( \Xp;\Yp,{Y}_{\mathsf{Fc}}|\Yc,\Xc \right) + I \left( \Xp,\Xc;\Yc \right),
   \label{eq:tuni1}
\\ \Rp + \Rc & \leq I \left( \Xc;\Yc|\Yp,\Xp \right) + I \left( \Xp,\Xc;\Yp \right).
   \label{eq:tuni2}
\end{align}
Notice that in the bounds in~\eqref{eq:cutset1 a}-\eqref{eq:tuni2}, ${Y}_{\mathsf{Fc}}$ always appears conditioned on $\Xc$. This implies that, for the ISD channel described in Section \ref{sec:ISDchannel}, ${Y}_{\mathsf{Fc}}$ can be replaced with $\Yf$ without loss of generality.

\item For the {\em memoryless ISD CCIC}, described in Section \ref{sec:ISDchannel}, with independent noises at the different source-destination pairs as in \eqref{eq:distr}, we have
\begin{align}
\Rp + \Rc
   \leq & I \left( \Yp;\Xp,\Xc|\Tp,\Yf \right) 
+  I \left( \Yc,\Yf; \Xp,\Xc|\Tc \right).
\label{eq:pv}
\end{align}
The details of the proof of the bound in~\eqref{eq:pv} can be found in Appendix \ref{app:eq:pv}. 

\end{subequations}

We note that a bound as the one in~\eqref{eq:pv} was originally derived in~\cite[Appendix IV pages 177-179]{PVIT11} for the ISD IC with bilateral source cooperation when  all noises are independent; for the case of unilateral source cooperation, this follows from the following Markov chain
\begin{subequations}
\begin{align}
&
(\Wp, \Xp^{i}) - (\Yf^{i-1}) - (\Wc, \Xc^{i},\Tc^{i}), \quad \forall i\in[1:N],
\label{eq:markovchain in PP1}
\\&
(\Wc, \Xc^{i}) - (\Yf^{i-1}) - (\Wp, \Xp^{i},\Tp^{i}), \quad \forall i\in[1:N].
\label{eq:markovchain in PP2}
\end{align}
\label{eq:markovchain in PP all}
\end{subequations}
A careful analysis of the bounding steps in \cite[Appendix IV pages 177-179]{PVIT11} shows that the derivation of the bound in~\eqref{eq:pv} is valid even when $\mathbb{P}_{{Y}_{\mathsf{Fc}},\Yc,\Yp|\Xp,\Xc}$ factors as in \eqref{eq:distr}, i.e., the independent noises assumption at all terminals captured by the product distribution $\mathbb{P}_{{Y}_{\mathsf{Fc}},\Yc,\Yp|\Xp,\Xc}=\mathbb{P}_{{Y}_{\mathsf{Fc}}|\Xp,\Xc} \mathbb{P}_{\Yc|\Xp,\Xc} \mathbb{P}_{\Yp|\Xp,\Xc}$ is not necessary for the bound to hold by suitably modifying the Markov chains in~\eqref{eq:markovchain in PP all} --- see Lemma~\ref{lemma2MC}. 
Among the advantages of the bound in~\eqref{eq:pv} is that the case of output feedback from the intended destination is a special case of the more general framework and can be obtained by ${Y}_{\mathsf{Fc}}=\Yc$.

We note that our bound in~\eqref{eq:pv} is not only more general but also tighter than the one in~\cite[Appendix IV pages 177-179]{PVIT11} 
since $I\left(\Yf; \Xp,\Xc| \Tc\right) \leq I \left( \Yf ; \Xp\right)$
and  $H \left( \Yp|\Tp,\Yf \right) \leq H \left( \Yp|\Tp\right)$.

The key step of the proof for the bound in~\eqref{eq:pv} is the following Lemma:
\begin{lemma}\label{lemma2MC}
For the ISD CCIC with the noise structure in \eqref{eq:distr}, the following Markov chains hold for all $i\in[1:N]$:
\begin{subequations}
\begin{align}
&(\Wp, \Tp^{i-1},\Xp^{i}) - (\Tc^{i-1}, \Yf^{i-1}) - (\Tc_{i}),          \label{eq:MC with Wp}\\
&(\Wc, \Tc^{i-1},\Xc^{i}) - (\Tp^{i-1}, \Yf^{i-1}) - (\Tp_{i}, \Yf_{i}). \label{eq:MC with Wc}
\end{align}
\label{eq:markovchain novel}
\end{subequations}
\end{lemma}
\begin{IEEEproof}
The proof 
is based on the Functional Dependence Graph (FDG) \cite{KramerPhD} and can be found in Appendix \ref{app:FDG}.
\end{IEEEproof}

\item For the {\em memoryless ISD IC with output feedback} ${Y}_{\mathsf{Fc}}=\Yc$ in \eqref{eq:distr}, from \cite[model-(1000)]{SahaiIT2013} we have
\begin{align}
\Rp + 2 \Rc  \leq & I \left( \Yc; \Xp,\Xc \right) + I \left( \Yc; \Xc|\Yp,\Xp \right) + I \left( \Yp; \Xp,\Xc|\Tp \right).
\label{eq:sahai}
\end{align}
To the best of our knowledge, the bound in~\eqref{eq:sahai} is the only upper bound of the type $\Rp + 2 \Rc$ available in the literature for the cooperative IC (which includes feedback models as a special case), but it is only valid for the case of output feedback.
Our goal in the next section is to derive bounds of the type of~\eqref{eq:sahai} for the class of ISD CCICs described in Section \ref{sec:ISDchannel} with independent noises at the different source-destination pairs as in \eqref{eq:distr}.

\end{enumerate}

\subsection{Novel outer bounds}
\label{sec:outerNew}
In this section we derive two novel outer bounds of the type $\Rp + 2 \Rc$ and $2 \Rp + \Rc$ on the capacity region for the ISD CCIC described in Section \ref{sec:ISDchannel} with independent noises at the different source-destination pairs as in \eqref{eq:distr}.  
These two outer bounds generalize to the CCIC those of the same type in \cite[Theorem 1]{TelatarTseISIT2007}, 
derived for the classical non-cooperative IC.

Our main result in this section is as follows.
\begin{theorem}
\label{th:main th 2r1+r2 and r1+2r2}
For the ISD CCIC satisfying the condition in~\eqref{eq:distr}, the capacity region is outer bounded by
\begin{align}
2 \Rp + \Rc
&\leq  I \left( \Yp;\Xp,\Xc \right) 
+ I \left( \Yp; \Xp|\Yf,\Yc,\Xc \right) 
+ I \left( \Yc,\Yf;\Xp,\Xc|\Tc \right),
\label{eq:bound2Rp+Rc}
\\
\Rp + 2\Rc
&\leq I \left( \Yc;\Xp,\Xc \right)
+I \left( \Yc;\Xc|\Yf,\Yp,\Xp \right) 
+I \left( \Yp,\Yf;\Xp,\Xc|\Tp \right).
\label{eq:boundRp+2Rc}
\end{align}
for some joint input distribution $\mathbb{P}_{\Xp,\Xc}$.
\end{theorem}

Note that, when evaluated for the case of output feedback with independent noises, i.e., $\Yf=\Tp$, the outer bound in \eqref{eq:boundRp+2Rc} reduces to the one in \eqref{eq:sahai}.

\begin{IEEEproof}
By Fano's inequality, by considering that the messages $\Wp$ and $\Wc$ are independent and by giving side information similarly to \cite{PVIT11}, we have
\begin{subequations}
\begin{align}
&N(2\Rp+\Rc-3\epsilon_N)
\notag\\ &\leq   
        2I \left( \Wp; \Yp^N \right)
      +  I \left( \Wc; \Yc^N \right) 
\notag\\ &\leq
          I \left( \Wp; \Yp^N \right)
    + I \left( \Wp; \Yp^N, \Tp^N, \Yf^N |\Wc\right)
   + I \left( \Wc; \Yc^N, \Tc^N, \Yf^N \right) 
\notag
\\& = H \left( \Yp^N \right )- H \left( \Yp^N, \Tp^N, \Yf^N |\Wp,\Wc \right)\label{eq:pair 1}
\\& + H \left( \Yp^N, \Tp^N, \Yf^N |\Wc \right) - H \left( \Yc^N, \Tc^N, \Yf^N|\Wc \right)\label{eq:pair 2}
\\& + H \left( \Yc^N, \Tc^N, \Yf^N \right) - H \left( \Yp^N| \Wp \right ).\label{eq:pair 3}
\end{align}
\end{subequations}
We now analyze and bound each pair of terms.
\paragraph*{Pair in~\eqref{eq:pair 1}}
We have
\begin{align*}
  &H \left( \Yp^N \right )- H \left( \Yp^N, \Tp^N, \Yf^N |\Wp,\Wc \right)
\\ \stackrel{({\rm a})}{=} & \sum_{i\in[1:N]} 
    H \left(\Yp_i |\Yp^{i-1}\right)
 - H \left(\Yp_{i}, \Tp_{i}, \Yf_{i} | \Wp,\Wc,\Yp^{i-1},\Tp^{i-1},\Yf^{i-1}, \Xp^{i}, \Xc^{i}\right)
\\ \stackrel{({\rm b})}{\leq} & \sum_{i\in[1:N]} 
    H \left(\Yp_i \right)
 - H \left(\Yp_{i}, \Tp_{i}, \Yf_{i} | \Wp,\Wc,\Yp^{i-1},\Tp^{i-1},\Yf^{i-1}, \Xp^{i}, \Xc^{i}\right)
\\ \stackrel{({\rm c})}{=} & \sum_{i\in[1:N]} 
    H \left(\Yp_i \right)
  - H \left(\Yp_{i}, \Tp_{i}, \Yf_{i} | \Xp_{i}, \Xc_{i}\right),
\end{align*}
where: the equality in $({\rm a})$ follows by applying the chain rule of the entropy and since, for the ISD CCIC, the encoding function $\Xc_{i}(\Wc,{Y}_{\mathsf{Fc}}^{i-1})$ is equivalent to $\Xc_{i}(\Wc,\Yf^{i-1})$ and since, given $\Wp$, $\Xp$ is uniquely determined;
the inequality in $({\rm b})$ is due to the conditioning reduces entropy principle; the equality in $({\rm c})$ follows because of the ISD property of the channel and since the channel is memoryless.

\paragraph*{Pair in~\eqref{eq:pair 2}}
We have
\begin{align*}
  &H \left( \Yp^N, \Tp^N, \Yf^N |\Wc \right)
 - H \left( \Yc^N, \Tc^N, \Yf^N|\Wc \right)
\\ \stackrel{({\rm d})}{=} & \sum_{i\in[1:N]}
   H \left( \Yp_{i}, \Tp_{i}, \Yf_{i}|\Yp^{i-1}, \Tp^{i-1},  \Yf^{i-1}, \Wc , \Xc^{i}\right)  -   H \left( \Yc_{i}, \Tc_{i}, \Yf_{i}|\Yc^{i-1}, \Tc^{i-1},  \Yf^{i-1}, \Wc , \Xc^{i}\right)
\\ \stackrel{({\rm e})}{=} & \sum_{i\in[1:N]}
   H \left( \Yp_{i}, \Tp_{i}, \Yf_{i}|\Yp^{i-1}, \Tp^{i-1},  \Yf^{i-1}, \Wc , \Xc^{i}\right)  -    H \left( \Tp_{i}, \Tc_{i}, \Yf_{i}|\Tp^{i-1}, \Tc^{i-1},  \Yf^{i-1}, \Wc , \Xc^{i}\right)
\\ \stackrel{({\rm f})}{\leq} & \sum_{i\in[1:N]}
   \underbrace{ H \left( \Tp_{i}, \Yf_{i}|           \Tp^{i-1},  \Yf^{i-1}, \Wc , \Xc^{i}\right)
 -    H \left( \Tp_{i}, \Yf_{i}|\Tp^{i-1}, \Tc^{i-1},  \Yf^{i-1}, \Wc , \Xc^{i}\right) }_{\text{$= 0$ because of~\eqref{eq:MC with Wc}}}
\\ &+ \sum_{i\in[1:N]}
   H \left( \Yp_i|\Tp_{i},  \Yf_{i}, \Xc_{i}\right)
-    H \left( \Tc_i|\Tp^{i}, \Tc^{i-1},  \Yf^{i}, \Wc , \Xc^{i}, \Xp^{i}\right)
\\ \stackrel{({\rm g})}{=} & \sum_{i\in[1:N]}
   H \left( \Yp_i|\Tp_{i},  \Yf_{i}, \Xc_{i}\right)
-  H \left( \Yp_i|\Tp_{i},  \Yf_{i}, \Xc_{i}, \Xp_i\right),
\end{align*}
where: the equality in $({\rm d})$ follows by applying the chain rule of the entropy and since, for the ISD CCIC, the encoding function $\Xc_{i}(\Wc,{Y}_{\mathsf{Fc}}^{i-1})$ is equivalent to $\Xc_{i}(\Wc,\Yf^{i-1})$; the equality in $({\rm e})$ is due to the fact that $\Yc$ is a deterministic function of $(\Xc,\Tp )$, which is invertible given $\Xc$; the inequality in $({\rm f})$ is due to the conditioning reduces entropy principle; the equality in $({\rm g})$ follows because of the ISD property of the channel and since the channel is memoryless.

\paragraph*{Pair in~\eqref{eq:pair 3}}
We have
\begin{align*}
& H \left ( \Yp^N|\Wp \right) 
 \\ \stackrel{({\rm h})}{=} &   \sum_{i\in[1:N]} H \left ( \Yp_i|\Yp^{i-1},\Wp,{\Xp}^i\right)
 \\ \stackrel{({\rm i})}{=} &  \sum_{i\in[1:N]} H \left ( \Tc_i|\Tc^{i-1},\Wp,{\Xp}^i\right)
\\ \stackrel{({\rm j})}{\geq} & \sum_{i\in[1:N]} H \left ( \Tc_i|\Tc^{i-1},\Wp,{\Xp}^i,\Yf^{i-1} \right)
\\ \stackrel{({\rm k})}{=} & \sum_{i\in[1:N]} 
	H \left ( \Tc_i|\Tc^{i-1},\Yf^{i-1} \right)
 - \underbrace{I \left( \Tc_i;\Wp,\Xp^{i}|\Tc^{i-1},\Yf^{i-1}\right)}_{\text{$=0$ because of~\eqref{eq:MC with Wp}}},
\end{align*}
where: the equality in $({\rm h})$ follows by applying the chain rule of the entropy and since, given $\Wp$, $\Xp$ is uniquely determined; the equality in $({\rm i})$ is due to the fact that $\Yp$ is a deterministic function of $(\Xp,\Tc )$, which is invertible given $\Xp$; the inequality in $({\rm j})$ follows since conditioning reduces the entropy; the equality in $({\rm k})$ follows from the definition of mutual information.
Therefore,
\begin{align*}
  &H \left( \Yc^N, \Tc^N, \Yf^N \right)
  -H \left( \Yp^N|\Wp \right)
\\ \stackrel{({\rm l})}{\leq} &\sum_{i\in[1:N]}
  H \left( \Yc_{i},\Tc_{i},\Yf_{i}|\Yc^{i-1},\Tc^{i-1},\Yf^{i-1}\right)
 - H \left( \Tc_i|\Tc^{i-1},\Yf^{i-1} \right)
\\ \stackrel{({\rm m})}{\leq} & \sum_{i\in[1:N]}
  H \left( \Tc_{i}|\Tc^{i-1},\Yf^{i-1}\right)
 - H \left( \Tc_i|\Tc^{i-1},\Yf^{i-1} \right) +
  H \left( \Yc_{i},\Yf_{i}|\Yc^{i-1},\Tc^{i},\Yf^{i-1}\right)
\\ \stackrel{({\rm n})}{\leq} & \sum_{i\in[1:N]} 0+
  H \left( \Yc_{i},\Yf_{i},|\Tc_i\right),
\end{align*}
where the inequality in $({\rm l})$ is a consequence of the inequality in $({\rm k})$ above and the inequalities in $({\rm m})$ and $({\rm n})$ are due to the conditioning reduces entropy principle.

\paragraph*{Final step}
By combining everything together, by introducing the time sharing random variable uniformly distributed over $[1:N]$ and independent of everything else, by dividing both sides by $N$ and taking the limit for $N\to\infty$ we get the bound in \eqref{eq:bound2Rp+Rc}. We finally notice that by dropping the time sharing we do not decrease the bound. Note also that, since for the ISD model defined in \eqref{eq:injsemidetMod} $\Tp$, respectively $\Tc$, is a deterministic function of $\left( \Yc,\Xc \right)$, respectively $\left( \Yp,\Xp \right)$, we have $H \left(\Tp,\Yf|\Yp,\Xp,\Xc  \right) = H \left(\Yc,\Yf|\Tc,\Xp,\Xc \right)$.

By following similar steps as in the derivation of \eqref{eq:bound2Rp+Rc} and by using the Markov chains in \eqref{eq:MC with Wp} and \eqref{eq:MC with Wc}, one can derive the upper bound in~\eqref{eq:boundRp+2Rc}. For completeness, we report the proof of~\eqref{eq:boundRp+2Rc} in Appendix \ref{app:proofboundRp+2Rc}.
\end{IEEEproof}

In the next section we will evaluate the outer bounds in \eqref{eq:outknownbefore} and those in Theorem \ref{th:main th 2r1+r2 and r1+2r2} for the Gaussian noise channel described in Section \ref{sec:ChGaussian} and show that they allow to characterize the capacity to within a constant gap in the regimes which were left open in \cite{ourITjournal}.

\section{The capacity region to within a constant gap for the symmetric GCCIC}
\label{sec:gap sumcapacity Symmetric Channel}

\begin{figure}
\centering
\includegraphics[width=0.65\textwidth]{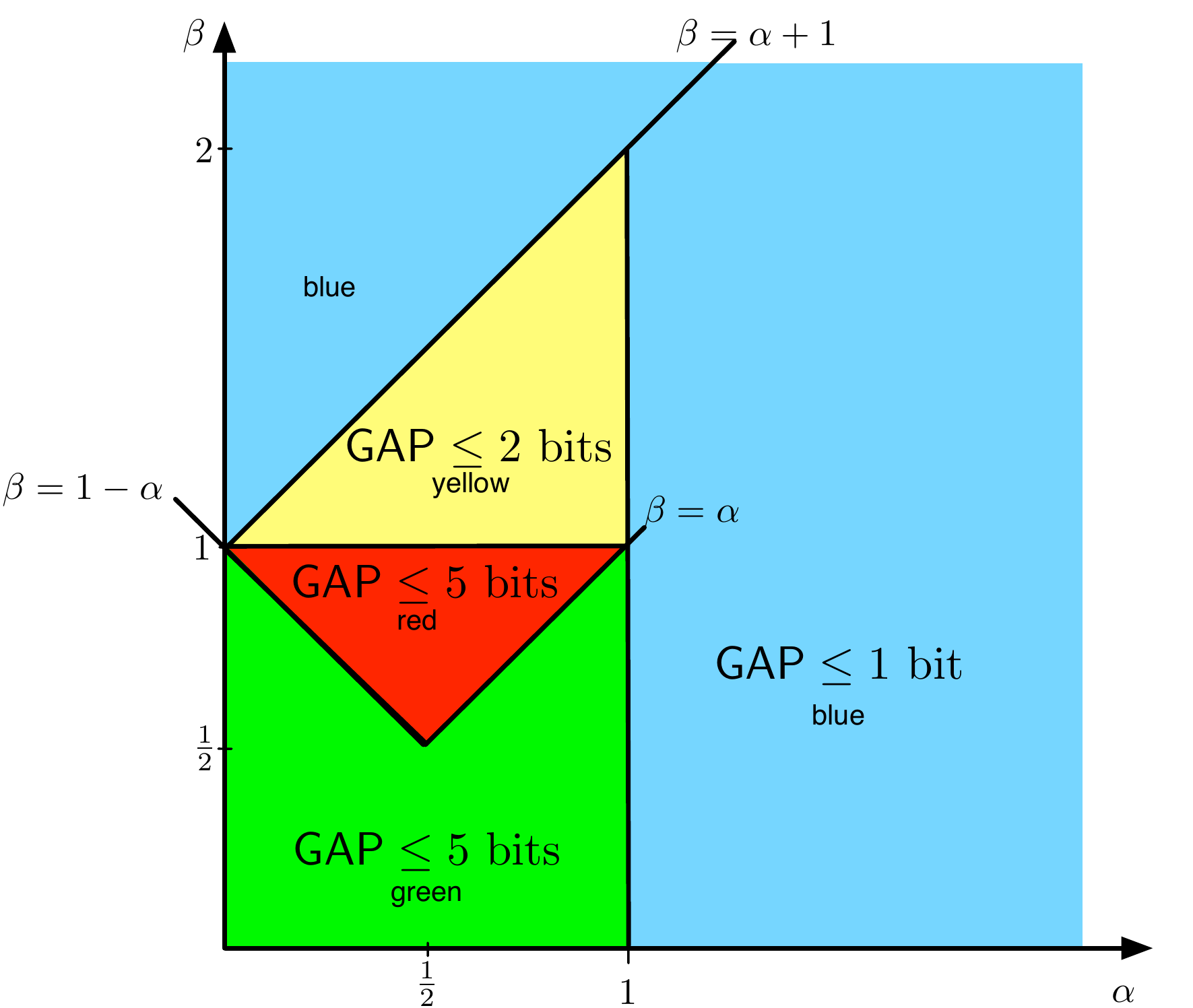}
\caption{Gap for different parameter regimes for the symmetric GCCIC. 
}
\label{fig:DoF}
\end{figure}

In this section we analyze the practically relevant Gaussian noise channel described in Section \ref{sec:ChGaussian}. In particular, we focus on the symmetric case defined in \eqref{eq:paramet}. 
For such a scenario, in \cite[Theorem 1]{ourITjournal}, we proved a constant gap of $1$ bit in the strong interference regime, i.e., $\inr \geq \snr$ (equivalent to $\alpha \geq 1$), and in the weak interference regime, i.e., $\inr < \snr$ (equivalent to $\alpha < 1$) when the cooperation link is `sufficiently strong' as quantified by $\Cc \geq \Delta_\text{th}$ with
\begin{align}
\Delta_\text{th}:=\left({\snr} + {\inr} + 2 \sqrt{{\inr}{\snr} \ \frac{{\inr}}{1+{\inr}}} \right)(1+{\inr}), \ \text{(equivalent to $\beta \geq \alpha +1$)}.
\label{eq:deltadef}
\end{align}
The regimes for which we proved the constant gap result of $1$ bit in \cite[Theorem 1]{ourITjournal} are depicted in blue in Fig. \ref{fig:DoF}, where the whole set of parameters has been partitioned into multiple sub-regions depending upon different levels of cooperation ($\beta$) and interference ($\alpha$) strengths.

Therefore, the case of weak interference ($\alpha < 1$) and weak cooperation ($\beta <\alpha+1$) (i.e., yellow, red and green regions in Fig. \ref{fig:DoF}) was left open in \cite{ourITjournal}, where
we speculated that for this regime, novel outer bounds of the type $2 \Rp+\Rc$ and $\Rp+2\Rc$ would be needed since, when the cooperation link is not `sufficiently strong', the performance of the GCCIC should reduce to the classical non-cooperative IC, whose capacity has bounds of the type $2 \Rp+\Rc$ and $\Rp+2\Rc$ \cite{etw}. With the two novel outer bounds in Theorem \ref{th:main th 2r1+r2 and r1+2r2} we can prove the main result of this section: 
\begin{theorem}
\label{thm:gap symmetric} 
The capacity region outer bound for the symmetric GCCIC in weak interference, i.e., ${\inr} < {\snr}$, is achievable to within a constant gap.
In particular:
\begin{enumerate}

\item 
$\Cc \leq \snr $ (i.e., $\beta \leq 1$), which corresponds to the green and red regions in Fig. \ref{fig:DoF}: $\mathsf{GAP} \leq 5$~bits.

\item 
$\snr < \Cc < \Delta_\text{th}$ (i.e., $1<\beta <1+\alpha$) for $\Delta_\text{th}$ in~\eqref{eq:deltadef}, which corresponds to the yellow region in Fig. \ref{fig:DoF}: $\mathsf{GAP} \leq 2$~bits.

\item 
For the remaining parameter regimes, which correspond to the blue region in Fig. \ref{fig:DoF}: $\mathsf{GAP} \leq 1$~bits~\cite{ourITjournal}.

\end{enumerate}
\end{theorem}

Theorem \ref{thm:gap symmetric} and \cite[Theorem 1]{ourITjournal} characterize the whole capacity region for the symmetric GCCIC to within $5$ bits. The rest of this section is dedicated to the proof of Theorem \ref{thm:gap symmetric}. In particular, in Section \ref{subsect:OBGaussian} we evaluate the outer bounds in \eqref{eq:outknownbefore} and those in Theorem \ref{th:main th 2r1+r2 and r1+2r2} for the symmetric GCCIC, in Section \ref{subsect:IBGaussian} we derive a novel achievable rate region and finally in Section \ref{subsec:gapGaussian} we show that the achievable rate region is a constant number of bits apart from the outer bound region.

\subsection{Outer bound region}
\label{subsect:OBGaussian}
We evaluate the bounds in 
\eqref{eq:outknownbefore},
\eqref{eq:bound2Rp+Rc} and \eqref{eq:boundRp+2Rc} for the Gaussian noise channel in \eqref{eq:awgn full}. We define  $\mathbb{E} \left [ \Xp \Xc^* \right ] := \rho: |\rho| \in [0,1]$. We also assume that all the noises are independent, which represents a particular case for which our outer bounds hold.
By the `Gaussian maximizes entropy' principle, jointly Gaussian inputs exhaust the outer bounds in \eqref{eq:outknownbefore}, \eqref{eq:bound2Rp+Rc} and \eqref{eq:boundRp+2Rc}.
Thus, we start by evaluating each mutual information term in \eqref{eq:outknownbefore}, \eqref{eq:bound2Rp+Rc} and \eqref{eq:boundRp+2Rc} by using jointly Gaussian inputs. Then, we further upper bound each mutual information term over the input correlation coefficient $\rho:|\rho| \in [0,1]$. By doing so we obtain:
\begin{lemma}
The capacity region of the symmetric GCCIC is contained into
\begin{subequations}
\begin{align}
   &\Rp  \leq  \log \left( 1+\Cc+\snr \right),
\label{eq:OBsymmetricGaussian1}
\\ &\Rp  \leq   \log\left ( 1+ \left (\sqrt{\snr} +  \sqrt{\inr} \right )^2\right ),
\label{eq:OBsymmetricGaussian2}
\\ &\Rc  \leq  \log \left( 1 +  \snr \right),
\label{eq:OBsymmetricGaussian3}
\\ & \Rp + \Rc  \leq \log  \left( 1 + \frac{ \snr}{1+ \inr}  \right) + \log  \left( 1+ \left (\sqrt{\snr} + \sqrt{\inr} \right)^2  \right),
\label{eq:OBsymmetricGaussian5}
\\ & \Rp + \Rc \leq \log  \underbrace{\left( 1 + \frac{ \snr +\Cc}{1+ \inr}  \right) +  \log  \left( 1+ \left(\sqrt{\snr} + \sqrt{\inr}\right)^2  \right)}_{\geq {\rm eq\eqref{eq:OBsymmetricGaussian5}}},
\label{eq:OBsymmetricGaussian4}
\\ & \Rp + \Rc \leq 
\underbrace{
\log \left( 1 + \Cc + \inr + \frac{\snr}{1+\inr} \right) +\log \left( 1 + \inr + \snr\frac{1+\Cc}{1+\Cc+\inr} \right) 
}_{
\stackrel{{\rm for} \ \Cc \geq \inr}{\geq}
\log \left( 1 + 2\inr + \frac{\snr}{1+\inr} \right) +\log \left( 1 + \inr + \frac{\snr}{2} \right)
\geq {\rm eq\eqref{eq:OBsymmetricGaussian5}} -2\log(2)
}
 + 2\log(2),
\label{eq:OBsymmetricGaussian6}
\\ & 2\Rp+\Rc \leq \underbrace{\log\left ( 1+\left (\sqrt{\snr} +  \sqrt{\inr} \right)^2\right ) 
+ \log \left( 1 + \frac{ \snr}{1+ \inr}  \right)}_{\text{\rm sum-rate bound in~\eqref{eq:OBsymmetricGaussian5}}} +\Delta_{\eqref{eq:OBsymmetricGaussian7}}  + \log(2), 
\label{eq:OBsymmetricGaussian7}
\\& \qquad  
\Delta_{\eqref{eq:OBsymmetricGaussian7}}:= 
\log \left( 1 \!+\! \frac{\Cc}{1+\inr+\snr} \right)\!+\!\log \left( 1 + \inr+\snr\frac{1+\Cc}{1+\inr+\Cc} \right)
\stackrel{{\rm for} \ \Cc \geq \inr}{\geq} \log \left (1+\inr+\frac{\snr}{2} \right),
\\ & \Rp + 2 \Rc \leq 
\underbrace{\log  \left( 1+ \left (\sqrt{\snr} + \sqrt{\inr} \right)^2  \right) + \log \left( 1 + \frac{ \snr}{1+ \inr}  \right)}_{\text{\rm sum-rate bound in~\eqref{eq:OBsymmetricGaussian5}}}  + \Delta_{\eqref{eq:OBsymmetricGaussian8}}+ \log(2),
\label{eq:OBsymmetricGaussian8}
\\ & \quad 
\Delta_{\eqref{eq:OBsymmetricGaussian8}}:= 
\log \left( 1 + \Cc+\inr+\frac{ \snr}{1+ \inr}  \right)\stackrel{{\rm for} \ \Cc \geq \snr}{\geq} \log(1+\snr).
\end{align}
\label{eq:OBsymmetricGaussian}
\end{subequations}
\end{lemma}
\begin{IEEEproof}
The outer bound region for the general GCCIC is given in~\eqref{eq:knownOBgeneralGaussian} in Appendix \ref{app:OBGaussianGeneral} that, specialized to the symmetric case, gives the region in~\eqref{eq:OBsymmetricGaussian}.

We note that the outer bound expressions in \eqref{eq:OBsymmetricGaussian} are slightly different from those reported in \cite[eq.(1)]{OurISCCSP2014} and \cite[eq.(6)]{OurISIT2014} because of different bounding steps over the input correlation factor $\rho, |\rho| \in [0,1]$;
the form presented in \eqref{eq:OBsymmetricGaussian}, although not being the tightest, is in our opinion the most amenable for easy closed-form gap computations.
\end{IEEEproof}


The following discussion applies `up to a constant gap', that is, by excluding terms that are not a function of the channel gains.
In weak interference the bound in~\eqref{eq:OBsymmetricGaussian1} can be dropped since the condition $\inr\leq\snr$  in~\eqref{eq:OBsymmetricGaussian2} implies $\Rp\leq \log\left( 1+\snr \right)+2\log(2)$. The bound in~\eqref{eq:OBsymmetricGaussian4} can also be dropped because it is looser than the one in~\eqref{eq:OBsymmetricGaussian5}. These observations imply that in the weak interference regime only the `ISD~bounds' in~\eqref{eq:OBsymmetricGaussian6}-\eqref{eq:OBsymmetricGaussian8} depend on  the strength of the cooperation link $\Cc$. 
The channel conditions for which the whole outer bound in~\eqref{eq:OBsymmetricGaussian} does not depend on $\Cc$ can be characterized as follows:

\begin{enumerate}

\item \label{rem:samePerfasIdeal}
$\Cc \geq \max\{\snr,\inr\}$: when the outer bound in~\eqref{eq:OBsymmetricGaussian} is an outer bound for the non-causal Gaussian CIC.
The bounds in~\eqref{eq:OBsymmetricGaussian7} and~\eqref{eq:OBsymmetricGaussian8} are redundant when $\Cc \geq \max\{\snr,\inr\}$. 
Moreover, for $\Cc \geq \inr$ the sum-rate in~\eqref{eq:OBsymmetricGaussian6} is looser than the one in~\eqref{eq:OBsymmetricGaussian5}.
We therefore conclude that 
for $\Cc \geq \max\{\snr,\inr\}$
the outer bound in~\eqref{eq:OBsymmetricGaussian} does not depend on $\Cc$ and reduces to the region 
\begin{subequations}
\begin{align}
   &\Rp  \leq  \log \left( 1+\left(\sqrt{\snr} +  \sqrt{\inr} \right)^2\right ),
\\ &\Rc  \leq  \log \left( 1 + \snr \right),
\\ &\Rp + \Rc  \leq \log\left( 1 + \frac{ \snr}{1+ \inr}  \right) 
            + \log  \left( 1+ \left (\sqrt{\snr} + \sqrt{\inr} \right)^2  \right), 
\end{align}
\end{subequations}
which is an outer bound on the capacity region of the non-causal Gaussian CIC (see~\cite[Theorem III.1]{riniJ1}). 
Thus, our outer bound predicts that, when $\Cc \geq \max\{\snr,\inr\}$ (yellow and part of the blue regions in Fig. \ref{fig:DoF}), the symmetric GCCIC should behave, up to constant gap, as the symmetric non-causal Gaussian CIC;
in Section \ref{susubsection:yellowregime}, we will formally prove this for the yellow regime in Fig. \ref{fig:DoF}, i.e., for $\inr\leq \snr \leq \Cc <\Delta_\text{th}$ with $\Delta_\text{th}$ defined in~\eqref{eq:deltadef} (the blue regimes  in Fig. \ref{fig:DoF}, i.e., $\snr\leq \inr$ and $\snr> \inr$ with $\Cc \geq \Delta_\text{th}$, were proved in \cite{ourITjournal}).




\item \label{rem:samePerfasClassicalIC}
$\Cc \leq \min\left\{\snr, \frac{\inr \left( 1+\inr\right)}{1+\snr} \right\}$: when the outer bound in~\eqref{eq:OBsymmetricGaussian} is an outer bound for the classical Gaussian IC.
In \cite{ourITjournal}, we proved this behavior in strong interference (i.e., $\snr \leq \inr$), in which case we have $\min\left\{\snr, \frac{\inr \left( 1+\inr\right)}{1+\snr} \right\} = \snr$. On the other hand, in weak interference we have $\min\left\{\snr, \frac{\inr \left( 1+\inr\right)}{1+\snr} \right\}=\frac{\inr \left( 1+\inr\right)}{1+\snr}$, in which case
the condition $\Cc \leq \frac{\inr \left( 1+\inr\right)}{1+\snr}$ implies $\Cc \leq \inr$ and $\snr \frac{1+\Cc}{1+\Cc +\inr} \leq \frac{\snr}{1+\inr} + \inr$; thus, the outer bound in~\eqref{eq:OBsymmetricGaussian} 
reduces to the region
\begin{subequations}
\begin{align}
   \Rp &\leq \log \left( 1 + \snr \right) + 2\log(2),
\\ \Rc & \leq \log \left( 1+\snr\right),
\\ \Rp + \Rc & \leq \log \left( 1 + \snr  \right) + \log \left( 1 + \frac{\snr}{1+\inr}\right) + 2\log(2),
\\ \Rp + \Rc & \leq \log \left( 1+ \inr + \frac{\snr}{1+\inr} \right)+\log \left( 1+ \inr + \frac{\snr}{1+\inr} \right)+4\log(2),
\\ 2 \Rp + \Rc & \leq \log \left( 1+\snr+\inr \right) + \log \left( \frac{1+\snr}{1+\inr}\right) + \log \left( 1+ \inr + \frac{\snr}{1+\inr} \right) + 5\log(2),
\\ \Rp + 2 \Rc & \leq \log \left( 1+\snr+\inr \right)+ \log \left( \frac{1+\snr}{1+\inr}\right) + \log \left( 1+ \inr + \frac{\snr}{1+\inr} \right) + 4\log(2),
\end{align}
\end{subequations}
which is an outer bound on the capacity region of the classical Gaussian IC (see~\cite[Theorem 3]{etw}). 
Thus, our outer bound predicts that, when $\Cc \leq \min\left\{\snr, \frac{\inr \left( 1+\inr\right)}{1+\snr} \right\}$, the symmetric GCCIC should behave, up to constant gap, as the non-cooperative Gaussian IC; we will formally prove this for the weak interference regime $\Cc \leq \frac{\inr \left( 1+\inr\right)}{1+\snr}, \inr \leq \snr$ in Section \ref{susubsection:greenregime} (the strong interference regime $\Cc \leq\snr\leq \inr$ was proved in \cite{ourITjournal}).

\end{enumerate}

\subsection{Transmission strategy and achievable rate region}
\label{subsect:IBGaussian}
We next design a transmission strategy for the CCIC.
In particular, since we aim to derive a scheme that is approximately optimal in weak interference, 
we consider both {\it private} and {\it common} messages for the PTx and the CTx, in the spirit of \cite{etw} for the classical IC. 
Moreover, depending on the strength of the cooperation link, the PTx might take advantage of the help of the CTx in transmitting its message, i.e., the messages of the PTx are both {\it cooperative} and {\it non-cooperative}. 

In order to enable cooperation, a block Markov coding scheme is used as follows. Transmission is over a frame of $B\gg 1$ slots. In slot $t\in[1:B]$, the PTx sends its old (cooperative common and private) messages $\left( W_{\mathsf{p}cc,t-1},W_{\mathsf{p}pc,t-1}\right) $ and superposes to them the new (cooperative common and private) messages $\left( W_{\mathsf{p}cc,t},W_{\mathsf{p}pc,t}\right)$ and the new non-cooperative common and private messages $\left( W_{\mathsf{p}cn,t},W_{\mathsf{p}pn,t}\right)$.
At the end of slot $t$, the CTx jointly decodes the new cooperative messages $\left( W_{\mathsf{p}cc,t},W_{\mathsf{p}pc,t}\right)$ after subtracting the contribution of the old messages $\left( W_{\mathsf{p}cc,t-1},W_{\mathsf{p}pc,t-1}\right)$. 
At the beginning of slot $t\in[1:B]$, the CTx knows the PTx's old private cooperative message $W_{\mathsf{p}pc,t-1}$ and, depending on the strength of the cooperation link compared to the interference link, the CTx might precode both its private and common non-cooperative messages $\left(W_{\mathsf{c}cn,t},W_{\mathsf{c}pn,t}\right)$ against the known interference, in such a way that the CRx does not experience interference from the codewords conveying these messages.
The destinations wait until the whole frame has been received and then proceed to simultaneous backward decode all messages.

The detailed derivation of the achievable rate region can be found in Appendix~\ref{sec:allachschemsappDPC}.
The achievable rate region in compact form can be obtained by applying the Fourier-Motzkin Elimination (FME) procedure on the rate constraints in \eqref{eq:DPC unified}.
Since the rate region in \eqref{eq:DPC unified} is specified by $8$ auxiliary random variables and by $20$ rate constraints, the FME turns out to be quite involved. 
However, depending on the strength of the cooperation link compared to the direct and interfering links, the PTx might not use some of the messages, i.e., the corresponding auxiliary random variables are set to a deterministic constant. In particular:
\begin{enumerate}
\item When $\beta \leq \max \left\{ \alpha, 1-\alpha \right \}$ (green region in Fig.~\ref{fig:DoF}), the cooperation link is quite weak;
we therefore expect the CCIC to `behave' as the classical non-cooperative IC \cite{etw} for which both {\it private} and {\it common non-cooperative} messages are approximately optimal. 
Differently from the classical IC, the PTx also conveys part of its message through the CTx. This {\it cooperative} message is {\it common}, and thus also decoded at the CRx. Actually, since the cooperation link is weak, the amount of information that can be decoded, and hence delivered, by the CTx is limited. Thus, there is no need to employ binning, i.e., the scheme is based on superposition only.
In other words, for this regime, the PTx does not make use of the private cooperative message, i.e., with reference to the transmission strategy in Appendix~\ref{sec:allachschemsappDPC}, we set $S_1 = Z_1 = \emptyset$. \label{item:greenregion}
\item When $\max \left\{ \alpha, 1-\alpha \right \} <\beta \leq 1$ (red region in Fig.~\ref{fig:DoF}), the cooperation link is stronger than the interfering link, but weaker than the direct link. 
Thus, on the one hand the PTx takes advantage of these channel conditions by using {\it cooperative} messages; on the other hand, the cooperation link is not strong enough (i.e., weaker than the direct link) to allow the CTx to fully decode the PTx's message, and hence the PTx also uses a {\it non-cooperative} message. In particular, the {\it non-cooperative} message of the PTx is {\it private}; this is because the interference is too weak and forcing the CRx to fully decode the PTx's message would constrain the rate too much. 
At the same time, the CTx can also benefit from the strength of the cooperation link and boost its rate performance by  precoding its message against the private cooperative message of the PTx, i.e., the scheme is based both on superposition and binning. In other words, for this regime, the PTx does not make use of the common non-cooperative message, i.e., with reference to the transmission strategy in Appendix~\ref{sec:allachschemsappDPC}, we set $U_1 = \emptyset$.
\label{item:redregion}
\item When $1 <\beta < \alpha + 1$ (yellow region in Fig.~\ref{fig:DoF}), the cooperation link is stronger than both the interfering and the direct links. Thus, the PTx takes advantage of the strong cooperation link and sends its message to the PRx with the help of the CTx, i.e., the messages of the PTx are only {\it cooperative}. Moreover, since the interference is weak, the messages of the CTx and of the PTx are both {\it common} and {\it private}. 
Also for this regime we use binning at the CTx. In other words, for this regime, the PTx does not make use of the non-cooperative messages, i.e., with reference to the transmission strategy in Appendix~\ref{sec:allachschemsappDPC}, we set $U_1 =T_1= \emptyset$.
\label{item:yellowregion}
\end{enumerate}
As per our discussion above, depending on the strength of the cooperation link compared to the direct and interfering links, some types of messages are not needed to achieve the outer bound to within a constant gap.
Thus, instead of performing the FME directly on the rate constraints in \eqref{eq:DPC unified}, we apply the FME for two special cases: 
when $S_1 = Z_1 = \emptyset$ (to obtain a scheme for the green region in Fig.~\ref{fig:DoF}) and
when $U_1 = \emptyset$ (to obtain a scheme for the red and yellow regions in Fig.~\ref{fig:DoF}; for the yellow region we then further set $T_1=\emptyset$). 
The details 
can be found in Appendix \ref{sub:noPCPTx} and Appendix \ref{sub:noCNCPTx}, respectively. In the next section, we will show that the two achievable rate regions 
achieve the outer bound in \eqref{eq:OBsymmetricGaussian} to within a constant gap.

\begin{remark}
Although in this work we focus on the symmetric GCCIC, i.e., the two direct and the two interfering links are of the same strength, the derived outer bound and the designed transmission strategy are valid for a general GCCIC, which is described by $5$ different channel gains. Extensions to the general case are 
part of future work.
\end{remark}

\begin{remark}
The novelty of the transmission strategy designed in Section \ref{subsect:IBGaussian} compared to those proposed in \cite{ourITjournal} lies in the fact that the PTx uses cooperative and non-cooperative messages together. In particular: (i) the scheme based on superposition coding in \cite[Appendix B]{ourITjournal} used $U_1=\emptyset$, i.e., the PTx does not use a non-cooperative common message, while the superposition based scheme here proposed has $U_1 \neq \emptyset$; (ii) the scheme based on binning and superposition coding in \cite[Appendix C]{ourITjournal} used $U_1=T_1=\emptyset$, i.e., the PTx does not use non-cooperative common and private messages, while the scheme here proposed has $U_1 \neq \emptyset$ and $T_1 \neq \emptyset$.
\end{remark}

\subsection{Constant gap characterization}
\label{subsec:gapGaussian} 
In the following we analyze the green, red and yellow regions in Fig.~\ref{fig:DoF} separately.

\subsubsection{Regime $\Cc \leq \max\left\{\inr,\frac{\snr}{1+\inr}\right\}, \inr\leq\snr$ (green region in  Fig.~\ref{fig:DoF})}
\label{susubsection:greenregime}
As remarked in item \ref{item:greenregion} in Section \ref{subsect:IBGaussian}, for this region we set $S_1=Z_1=\emptyset$ in the transmission strategy in Appendix~\ref{sec:allachschemsappDPC}, i.e., the PTx does not use private cooperative messages. With this choice and after performing the FME, we obtain the rate region in \eqref{eq:noPCPTx} in Appendix \ref{sub:noPCPTx}, which evaluated for the Gaussian noise case gives the region in~\eqref{eq:noPCPTxGaussian}. In \eqref{eq:noPCPTxGaussian}, we set $\Ip=\Ic=\inr$, $\Sp=\Sc=\snr$ (i.e., we consider the symmetric case) and 
$|b_2|^2= 1-|a_2|^2 = \frac{1}{1+\inr}$ so that the private message of CTx (conveyed by $T_2$) is received below the noise level at the PRx in the spirit of \cite{etw}. Regarding the choice of the power splits for the PTx, we further split the green region into two subregions: subregion (i) for which $\Cc \leq \frac{\inr \left( 1+\inr\right)}{1+\snr}$ (i.e., $\beta \leq [2\alpha -1]^+$) and subregion (ii) for which $\Cc > \frac{\inr \left( 1+\inr\right)}{1+\snr}$ (i.e., $\beta > [2\alpha -1]^+$). We now analyze these two subregions separately.

{\it Subregion (i)}: when $\beta \leq [2\alpha -1]^+$, the cooperation link is very weak and thus we expect the GCCIC to behave as the non-cooperative Gaussian IC \cite{etw}. Therefore, we set the power  of the cooperative common message (carried by $V_1$) to $|b_1|^2=0$ in \eqref{eq:noPCPTxGaussian} and 
$|c_1|^2 = 1-|a_1|^2=\frac{1}{1+\inr}$ so that the private message of PTx (conveyed by $T_1$) is received below the noise level at the CRx in the spirit of \cite{etw}.
With these choices and by removing the redundant constraints in \eqref{eq:noPCPTxGaussian} (i.e., eq\eqref{eq:c1C1Gaussian}, eq\eqref{eq:c5C1Gaussian}, eq\eqref{eq:c6C1Gaussian}, eq\eqref{eq:c7C1Gaussian}, eq\eqref{eq:c10C1Gaussian} and eq\eqref{eq:c11C1Gaussian}), we get that the achievable rate region in \eqref{eq:noPCPTxGaussian} is contained into
\begin{subequations}
\label{eq:lowGreeni}
\begin{align}
\mathcal{I}^{\text{green-(i)}} : \quad \Rp &\leq \log \left( 1+\snr \right) - \log(2), \label{eq:lowGreeniA}
\\ \Rc &\leq \log \left( 1+\snr \right) - \log(2), \label{eq:lowGreeniB}
\\ \Rp + \Rc & \leq \log \left( 1+\snr + \inr\right) + \log \left( 1+\frac{\snr}{1+\inr} \right) - 2 \log(2), \label{eq:lowGreeniC}
\\ \Rp + \Rc & \leq \log \left( 1+ \inr + \frac{\snr}{1+\inr} \right) + \log \left( 1+ \inr + \frac{\snr}{1+\inr} \right)- 2 \log(2), \label{eq:lowGreeniD}
\\ 2 \Rp + \Rc & \leq \log \left( 1+\frac{\snr}{1+\inr}\right) + \log \left( 1+\snr +\inr\right) + \log \left( 1+ \inr + \frac{\snr}{1+\inr} \right) - 3 \log(2), \label{eq:lowGreeniE}
\\ \Rp+2\Rc & \leq  \log \left( 1+ \inr + \frac{\snr}{1+\inr} \right) + \log \left( 1+\frac{\snr}{1+\inr}\right) + \log \left( 1+\snr +\inr\right) - 3 \log(2). \label{eq:lowGreeniF}
\end{align}
\label{eq:lowGreeniALL}
\end{subequations}
Notice that the rate region in~\eqref{eq:lowGreeniALL} is the achievable region for the classical symmetric non-cooperative IC in weak interference, which is optimal up to a gap of 1~bit/user~\cite{etw}.
 
For this regime, the outer bound in \eqref{eq:OBsymmetricGaussian} can be further upper bounded (by removing the constraints in \eqref{eq:OBsymmetricGaussian2} and \eqref{eq:OBsymmetricGaussian4}) as
{
\begin{subequations}
\label{eq:outGreeni}
\begin{align}
\mathcal{O}^{\text{green-(i)}} : \quad
\Rp  & 
\leq \log \left( 1+\snr \right) + \log(2), \label{eq:outGreeniA}
\\ \Rc &  \leq  \log \left( 1 +  \snr \right), \label{eq:outGreeniB}
\\   \Rp + \Rc  & 
\leq \log  \left( 1 + \frac{ \snr}{1+ \inr}  \right) + \log  \left( 1+ \snr+ \inr\right) +  \log(2), \label{eq:outGreeniC}
\\ \Rp + \Rc & 
%
\stackrel{0\leq\Cc \leq \inr}{\leq} \log \left( 1  + \inr + \frac{\snr}{1+\inr} \right) +\log \left( 1 + \inr + \snr\frac{1+\Cc}{1+\inr} \right) +3\log(2) \nonumber
\\& \stackrel{\Cc \leq \frac{\inr \left( 1+\inr\right)}{1+\snr}}{\leq} \log \left( 1+ \inr + \frac{\snr}{1+\inr} \right) + \log \left( 1+ \inr + \frac{\snr}{1+\inr} \right) + 4 \log(2), \label{eq:outGreeniD}
\\ 2\Rp+\Rc & 
%
\stackrel{ 0\leq\Cc \leq \inr}{\leq} \log \left( 1+\snr +\inr\right)+ \log \left( 1 + \frac{ \snr}{1+ \inr}  \right) 
 + \log \left( 1 + \inr+\snr\frac{1+\Cc}{1+\inr} \right) + 3 \log(2)\nonumber
\\& \stackrel{\Cc \leq \frac{\inr \left( 1+\inr\right)}{1+\snr}}{\leq} 
 \log \left( 1 + \snr  + \inr\right) +  \log\left( 1 + \frac{\snr}{1 + \inr}\right)  +  \log \left( 1 +  \inr  +  \frac{\snr}{1 + \inr} \right)  +  4 \log(2), \label{eq:outGreeniE}
\\ \Rp+2\Rc & 
%
\stackrel{0\leq\Cc \leq \inr}{\leq} \log \left( 1 + \snr  + \inr\right)  +  \log \left( 1 + \frac{\snr}{1 + \inr}\right)   + \log \left( 1 +  \inr  +  \frac{\snr}{1+\inr} \right)  +  3 \log(2). \label{eq:outGreeniF}
\end{align}
\end{subequations}
}

It is easy to see that the outer bound region in \eqref{eq:outGreeni} and the achievable rate region in \eqref{eq:lowGreeni} are to within $3$ bits/user of one another. 
Notice that in order to prove a constant gap we compared: 
eq\eqref{eq:lowGreeniA} with eq\eqref{eq:outGreeniA},
eq\eqref{eq:lowGreeniB} with eq\eqref{eq:outGreeniB},
eq\eqref{eq:lowGreeniC} with eq\eqref{eq:outGreeniC},
eq\eqref{eq:lowGreeniD} with eq\eqref{eq:outGreeniD},
eq\eqref{eq:lowGreeniE} with eq\eqref{eq:outGreeniE},
and eq\eqref{eq:lowGreeniF} with eq\eqref{eq:outGreeniF}.

{\it Subregion (ii)}: when $\beta > [2\alpha -1]^+$, the GCCIC starts to benefit from cooperation and indeed the outer bound region depends on $\Cc$. Therefore the cooperative common message carried by $V_1$ can boost the rate performance of the system.
In \eqref{eq:noPCPTxGaussian}, we set the power of the common non-cooperative message (carried by $U_1$) to
$|a_1|^2= \frac{1}{2\left(1+\min\left\{\Cc,\inr \right \}\right)}$.
This choice is motivated by the fact that, in order to approximately match the outer bound, the single rate constraint on $\Rp$ in \eqref{eq:c2C1Gaussian} must approximately behave as an interference-free point-to-point channel. 
Therefore, the fact that the CTx can now decode part of the message of the PTx (carried by $V_1$) must not limit (up to a constant gap) the performance of the PTx. 
In other words, since $\Cc$ is `quite large' but not `huge', the rate of $V_1$ cannot be too large.
Moreover, we set $|c_1|^2 = \frac{1}{2(1+\inr)}$ so that the private message of PTx (conveyed by $T_1$) is received below the noise level at the CRx in the spirit of \cite{etw}. Thus, if $\inr \leq \Cc$ we have $|b_1|^2 = \frac{\inr}{1+\inr}$, while if $\inr > \Cc$ we have $|b_1|^2 = \frac{\Cc + \inr + 2 \Cc \inr}{2 (1+ \Cc) (1+\inr)}$.
With these choices and by removing the redundant constraints in \eqref{eq:noPCPTxGaussian} (i.e., 
eq\eqref{eq:c1C1Gaussian}, 
eq\eqref{eq:c4C1Gaussian}, 
eq\eqref{eq:c6C1Gaussian}, 
eq\eqref{eq:c7C1Gaussian}, 
and 
eq\eqref{eq:c12C1Gaussian}), 
we get that the achievable rate region in \eqref{eq:noPCPTxGaussian} is contained into
\begin{subequations}
\label{eq:lowGreenii}
\begin{align}
\mathcal{I}^{\text{green-(ii)}} : \quad \Rp &\leq \log \left( 1+\snr \right) - 4 \log(2), \label{eq:lowGreeniiA}
\\ \Rc &\leq \log \left( 1+\snr \right) - \log(2), \label{eq:lowGreeniiB}
\\ \Rp + \Rc & \leq \log \left( 1+\snr + \inr\right) + \log \left( 1+\frac{\snr}{1+\inr} \right) - 3 \log(2), \label{eq:lowGreeniiC}
\\ \Rp + \Rc & \leq \log \left( 1 +  \inr  +  \frac{\snr}{1+\inr} \right)  +  \log \left( 1 +  \frac{\snr}{1 + \inr} \right)
\notag\\& \quad +   \log \left( 1 + \min \left \{ \inr,\Cc\right\}\right) -  5 \log(2),
\label{eq:lowGreeniiD}
\\ 2 \Rp + \Rc & \leq 2 \log \left( 1+\frac{\snr}{1+\inr}\right)  +  \log \left( 1+\snr +\inr\right) 
\notag\\& \quad  +  \log \left ( 1 + \min \{\inr, \Cc \}\right )  -  6 \log(2),
\label{eq:lowGreeniiE}
\\ 2 \Rp + \Rc & \leq 2\log \left( 1+\frac{\snr}{1+\inr}\right) + \log \left( 1 + \inr + \frac{\snr}{1+\min \{\inr,\Cc\}}\right) 
\notag\\& \quad + 2\log \left ( 1 + \min \{\inr, \Cc \}\right ) - 9 \log(2),
\label{eq:lowGreeniiF}
\\ \Rp+2\Rc & \leq  \log \left( 1+ \inr + \frac{\snr}{1+\inr} \right) + \log \left( 1+\frac{\snr}{1+\inr}\right)
\notag\\& \quad  + \log \left( 1+\snr +\inr\right) - 4 \log(2). \label{eq:lowGreeniiG}
\end{align}
\end{subequations}

For this regime, the outer bound in \eqref{eq:OBsymmetricGaussian} can be further upper bounded (by removing the constraints in \eqref{eq:OBsymmetricGaussian2} and \eqref{eq:OBsymmetricGaussian4}) as
{
\begin{subequations}
\label{eq:outGreenii}
\begin{align}
\mathcal{O}^{\text{green-(ii)}} : \quad \Rp  & \leq  \log \left( 1+\snr \right) + \log(2),
\label{eq:outGreeniiA}
\\ \Rc &  \leq  \log \left( 1 +  \snr \right), \label{eq:outGreeniiB}
\\ \Rp + \Rc  & \leq   \log  \left( 1+ \snr + \inr  \right) + \log  \left( 1 + \frac{ \snr}{1+ \inr}  \right) + \log(2),
\label{eq:outGreeniiC}
\\ \Rp + \Rc & \stackrel{0 \leq \Cc \leq \max \left \{ \inr, \frac{\snr}{1+\inr}\right \}}{\leq} \log \left( 1  +  \inr  +  \frac{\snr}{1 + \inr} \right)
\notag\\& \quad   +  \log \left( 1  +  \inr +  \frac{\snr\left( 1 + \Cc\right) }{1 + \inr + \Cc} \right) + 3\log(2),
\label{eq:outGreeniiD}
\\ 2 \Rp + \Rc & \stackrel{0 \leq \Cc \leq \snr}{\leq}   \log \left( 1 + \frac{\snr}{1+\inr} \right) + \log\left ( 1+\snr +  \inr\right ) 
\notag\\& \quad   + \log \left( 1  + \inr+ \frac{\snr\left( 1+\Cc\right)}{1+\Cc+\inr} \right)+3\log(2),
\label{eq:outGreeniiE}
\\ \Rp + 2 \Rc & \stackrel{0 \leq \Cc \leq \max \left \{ \inr, \frac{\snr}{1+\inr}\right \}}{\leq} \log \left( 1+ \inr + \frac{\snr}{1+\inr} \right)+ \log \left( 1+\frac{\snr}{1+\inr}\right) \nonumber
\\& \quad + \log \left( 1+\snr +\inr\right) +3\log(2). \label{eq:outGreeniiF}
\end{align}
\end{subequations}
}

It is easy to see that the outer bound region in \eqref{eq:outGreenii} and the achievable rate region in \eqref{eq:lowGreenii} are to within $5$ bits/user of one another. 
Notice that in order to prove a constant gap we compared:
eq\eqref{eq:lowGreeniiA} with eq\eqref{eq:outGreeniiA},
eq\eqref{eq:lowGreeniiB} with eq\eqref{eq:outGreeniiB},
eq\eqref{eq:lowGreeniiC} with eq\eqref{eq:outGreeniiC},
eq\eqref{eq:lowGreeniiD} with eq\eqref{eq:outGreeniiD},
eq\eqref{eq:lowGreeniiE} with eq\eqref{eq:outGreeniiE},
eq\eqref{eq:lowGreeniiF} with eq\eqref{eq:outGreeniiE},
and eq\eqref{eq:lowGreeniiG} with eq\eqref{eq:outGreeniiF}.

\subsubsection{Regime $ \max\left\{\inr,\frac{\snr}{1+\inr}\right\} < \Cc \leq \snr$ (red region in  Fig.~\ref{fig:DoF})}
As remarked in item \ref{item:redregion} in Section \ref{subsect:IBGaussian}, for this region we set $U_1=\emptyset$ in the transmission strategy in Appendix~\ref{sec:allachschemsappDPC}, i.e., the PTx does not use a common non-cooperative message. With this choice and after performing the FME, we obtain the rate region in \eqref{eq:noCNCPTx} in Appendix \ref{sub:noCNCPTx}, which evaluated for the Gaussian noise case gives the region in~\eqref{eq:noCNCPTxGaussian}. For $\Ip=\Ic=\inr$ and $\Sp=\Sc=\snr$, we set $|a_2|=0$, 
and $|c_2|^2=1-|b_2|^2= \frac{1}{1+\inr}$ in the region in \eqref{eq:noCNCPTxGaussian}. With this choice the private message of the CTx (conveyed by $T_2$) is received below the noise level at the PRx in the spirit of \cite{etw}. Note also that the CTx does not cooperate with the PTx in conveying information to the PRx, but it just exploits the information it learns through the cooperation link to smartly pre-encode its messages. 
For the PTx we let $|a_1|^2 = |b_1|^2 = \frac{\inr + \Cc + 2 \inr \Cc}{4(1+\inr) (1+\Cc)}$, $|c_1|^2 = \frac{1}{2(1+\inr)}$ and $|d_1|^2 = \frac{1}{2(1+\Cc)}$; with this choice of the power splits and since we are in the regime $\Cc > \inr$, the two private messages of the PTx (i.e., the cooperative one carried by $Z_1$ and the non-cooperative one carried by $T_1$) are received at most at the level of the noise at the CRx. Moreover, the non-cooperative private message (carried by $T_1$) is received at the level of the noise at the CTx.

With these choices we get that the achievable rate region in \eqref{eq:noCNCPTxGaussian} is contained into 
(by considering $\min \left \{ k_1,k_2\right \} \geq 0$ in \eqref{eq:noCNCPTxGaussian})
\begin{subequations}
\label{eq:lowRed}
\begin{align}
\mathcal{I}^{\text{red}} : \quad \Rp &\leq \log \left( 1 + \Cc + \snr \right) - 5 \log(2), \label{eq:lowRedA}
\\ \Rp & \leq \log \left( 1 + \snr + \inr \right) - \log(2), \label{eq:lowRedB}
\\ \Rc & \leq \log \left( 1+\snr \right) - \log(3), \label{eq:lowRedC}
\\ \Rp + \Rc & \leq \log \left( 1+ \Cc \right)  +  \log \left ( 1 + \frac{\snr}{1+ \Cc} + \inr \right) 
\nonumber
 \\& \qquad +  \log \left( 1 + \frac{\snr}{1+\inr}\right)  -  5 \log(2)  -  \log(3), \label{eq:lowRedD}
\\ \Rp + \Rc & \leq \log \left( 1+\snr +\inr \right) + \log \left( 1 + \frac{\snr}{1+\inr}\right)-\log(2)-\log(3), \label{eq:lowRedE}
\\ \Rp + \Rc & \leq \log \left( 1 + \frac{\Cc}{1+\inr} \right) + \log \left(  1 + \frac{\snr}{1+\Cc}\right) + \log \left( 1+\snr\right)-4\log(2)-\log(3),\label{eq:lowRedF}
\\ \Rp + \Rc & \leq \log \left( 1 + \snr \right) + \log \left( 1 + \frac{\snr}{1+\inr} + \frac{\snr}{1+\Cc}\right) - 2 \log(2) - \log(3), \label{eq:lowRedG}
\\ \Rp + 2 \Rc & \leq \log \left( 1+\snr +\inr \right) + \log \left( 1+\frac{\snr}{1+\inr}\right) + \log \left( 1 + \snr\right) - 3 \log(2) -2\log(3), \label{eq:lowRedH}
\\ \Rp + 2 \Rc & \leq \log \left( 1+ \snr \right) + \log \left( 1 + \frac{\snr}{1+\inr}\right) + \log \left( 1+\inr + \frac{\snr}{1+\Cc} \right)\nonumber
 \\& \qquad + \log \left( 1+\frac{\Cc}{1+\inr}\right) -4 \log(2) - 2\log(3), \label{eq:lowRedI}
\\ \Rp + 3 \Rc & \leq \log \left( 1+\snr + \inr\right) + 2 \log \left(1+\frac{\snr}{1+\inr} \right) + \log \left( 1 + \inr + \frac{\snr}{1+\Cc} \right) \nonumber
\\& \qquad \log \left( 1 + \snr\right)-5 \log(2)-3 \log(3). \label{eq:lowRedL}
\end{align}
\end{subequations}

For this regime, the outer bound in \eqref{eq:OBsymmetricGaussian} can be further upper bounded (by considering the constraints in \eqref{eq:OBsymmetricGaussian1}, \eqref{eq:OBsymmetricGaussian2}, \eqref{eq:OBsymmetricGaussian3}, \eqref{eq:OBsymmetricGaussian5} and \eqref{eq:OBsymmetricGaussian8})
{
\begin{subequations}
\label{eq:outRed}
\begin{align}
\mathcal{O}^{\text{red}} : \quad \Rp  & \leq \log \left( 1 + \Cc + \snr \right), \label{eq:outRedA}
\\ \Rp &
\leq \log \left( 1 + \snr + \inr \right) + \log(2), \label{eq:outRedB}
\\ \Rc & \leq \log \left( 1+ \snr \right), \label{eq:outRedC}
\\ \Rp + \Rc & 
\leq \log \left( 1+\frac{\snr}{1+\inr} \right)+ \log \left( 1+\snr+\inr \right)+\log(2),
 \label{eq:outRedD}
\\ \Rp + 2 \Rc & 
\leq \log \left( 1+\snr+\inr \right) +\log \left( 1+\frac{\snr}{1+\inr} \right)+ \log \left( 1  + \Cc \right) +2\log(2)+\log(3),
\label{eq:outRedE}
\end{align}
\end{subequations}
where the inequality in \eqref{eq:outRedE} follows since $\inr + \frac{\snr}{1+\inr} \leq 2 \max \left \{ \inr, \frac{\snr}{1+\inr}\right \} \leq 2 \Cc$.}

It is easy to see that the outer bound region in \eqref{eq:outRed} and the achievable rate region in \eqref{eq:lowRed} are to 
within $5$ bits/user of one another. 
Notice that in order to prove a constant gap we compared: eq\eqref{eq:lowRedA} with eq\eqref{eq:outRedA},
eq\eqref{eq:lowRedB} with eq\eqref{eq:outRedB},
eq\eqref{eq:lowRedC} with eq\eqref{eq:outRedC},
eq\eqref{eq:lowRedD} with eq\eqref{eq:outRedD},
eq\eqref{eq:lowRedE} with eq\eqref{eq:outRedD},
eq\eqref{eq:lowRedF} with eq\eqref{eq:outRedD},
eq\eqref{eq:lowRedG} with eq\eqref{eq:outRedD},
eq\eqref{eq:lowRedH} with eq\eqref{eq:outRedC}+eq\eqref{eq:outRedD},
eq\eqref{eq:lowRedI} with eq\eqref{eq:outRedE},
and eq\eqref{eq:lowRedL} with eq\eqref{eq:outRedC}+eq\eqref{eq:outRedE}.

\subsubsection{Regime $\inr\leq \snr<\Cc < \Delta_\text{th}$ with $\Delta_\text{th}$ defined in \eqref{eq:deltadef} (yellow region in  Fig.~\ref{fig:DoF})}
\label{susubsection:yellowregime}
As remarked in item \ref{item:yellowregion} in Section \ref{subsect:IBGaussian}, for this region we set $U_1=T_1=\emptyset$ in the transmission strategy in Appendix~\ref{sec:allachschemsappDPC}, i.e., the PTx uses only cooperative messages. With this choice and after performing the FME, we obtain the rate region in \eqref{eq:noCNCPTx} in Appendix \ref{sub:noCNCPTx}, which evaluated for the Gaussian noise case gives the region in~\eqref{eq:noCNCPTxGaussian}. 
In \eqref{eq:noCNCPTxGaussian}, for $\Ip=\Ic=\inr$ and $\Sp=\Sc=\snr$, we set $|a_1|^2=|b_1|^2=\frac{\inr}{2 \left( 1+\inr\right)}$, $|b_2|^2=\frac{\inr}{1+\inr}$ and $|c_1|^2 = |c_2|^2 = \frac{1}{1+\inr}$. 
Notice that with this choice of the power splits, we have $|d_1|=0$, i.e., the power allocated for $T_1$ is zero.
With these choices, we get that the achievable rate region in \eqref{eq:noCNCPTxGaussian} can be further lower bounded (by considering $\min \left \{ k_1,k_2\right \} \geq 0$ and that the constraint in \eqref{eq:c5C2Gaussian} is redundant in \eqref{eq:noCNCPTxGaussian}) as
\begin{subequations}
\label{eq:lowYellow}
\begin{align}
\mathcal{I}^{\text{yellow}} : \quad \Rp &\leq \log \left( 1+\Cc \right) - \log(2), \label{eq:lowYellowA}
\\ \Rp & \leq \log \left( 1+\snr+\inr \right) - \log(2), \label{eq:lowYellowB}
\\ \Rc & \leq \log \left( 1 + \snr \right) - \log(2),
\label{eq:lowYellowC}
\\ \Rp + \Rc & \leq \log \left( 1+\Cc \right) + \log \left( 1+\snr+\inr\right)-3\log(2),
\label{eq:lowYellowD}
%
%
\\ \Rp + \Rc & \leq \log \left( 1+\frac{\Cc}{1+\inr} \right) + \log \left(1+\snr \right) - \log(2), \label{eq:lowYellowE}
\\ \Rp + \Rc & \leq \log \left( 1+\frac{\snr}{1+\inr} \right) + \log \left(1+\snr \right) - 2\log(2), \label{eq:lowYellowF}
\\ \Rp + 2 \Rc & \leq \log \left( 1+\snr \right)  + \log \left( 1+\frac{\snr}{1+\inr} \right) + \log \left(1+\snr+\inr \right) - 4 \log(2),
\label{eq:lowYellowG}
\\ \Rp + 2 \Rc & \leq \log \left( 1+\frac{\Cc}{1+\inr} \right) + \log \left(1+\snr+\inr \right)+ \log \left(1+\snr \right) -3 \log(2),
\label{eq:lowYellowH}
\\ \Rp + 3 \Rc & \leq 2\log \left(1+\snr+\inr \right)+ \log \left(1+\snr \right) + \log \left( 1+\frac{\snr}{1+\inr} \right) -6 \log(2).
\label{eq:lowYellowI}
\end{align}
\end{subequations}

For this regime, the outer bound in \eqref{eq:OBsymmetricGaussian} can be further upper bounded (by considering the constraints in \eqref{eq:OBsymmetricGaussian1}, \eqref{eq:OBsymmetricGaussian2}, \eqref{eq:OBsymmetricGaussian3}, and \eqref{eq:OBsymmetricGaussian5})
{
\begin{subequations}
\label{eq:outYellow}
\begin{align}
\mathcal{O}^{\text{yellow}} : \quad \Rp  &
\leq \log \left( 1 + \Cc \right) + \log(2), \label{eq:outYellowA}
\\ \Rp &
\leq \log \left( 1 + \snr + \inr \right) + \log(2), \label{eq:outYellowB}
\\ \Rc & \leq \log \left( 1+\snr \right),
\label{eq:outYellowC}
%
%
\\ \Rp + \Rc & 
\leq \log \left( 1+\frac{\snr}{1+\inr} \right)+ \log \left( 1+\snr+\inr \right)+\log(2).
\label{eq:outYellowE}
\end{align}
\end{subequations}
}

It is easy to see that the outer bound region in \eqref{eq:outYellow} and the achievable rate region in \eqref{eq:lowYellow} are to within $2$ bits/user of one another. Notice that in order to prove a constant gap we compared: 
eq\eqref{eq:lowYellowA} with eq\eqref{eq:outYellowA},
eq\eqref{eq:lowYellowB} with eq\eqref{eq:outYellowB},
eq\eqref{eq:lowYellowC} with eq\eqref{eq:outYellowC},
eq\eqref{eq:lowYellowD} with eq\eqref{eq:outYellowA}+eq\eqref{eq:outYellowC},
eq\eqref{eq:lowYellowE} with eq\eqref{eq:outYellowE},
eq\eqref{eq:lowYellowF} with eq\eqref{eq:outYellowE},
eq\eqref{eq:lowYellowG} with eq\eqref{eq:outYellowC}+eq\eqref{eq:outYellowE},
eq\eqref{eq:lowYellowH} with eq\eqref{eq:outYellowC}+eq\eqref{eq:outYellowE},
and eq\eqref{eq:lowYellowI} with 2eq\eqref{eq:outYellowC}+eq\eqref{eq:outYellowE}.

\subsection{Discussion} 

The two novel outer bounds $2\Rp+\Rc$ in \eqref{eq:OBsymmetricGaussian7} and $\Rp+2\Rc$ in \eqref{eq:OBsymmetricGaussian8} are active when $\snr \geq \max \left \{ \Cc, \inr\right \}$ (weak interference and weak cooperation, which corresponds to the red and green regions in Fig.~\ref{fig:DoF}). 
In \cite{suhtse:ICwithfeedback}, the authors interpreted the need of this type of bounds as a measure of the amount of the `resource holes', or inefficiency, due to the distributed nature of the non-cooperative classical IC \cite{etw}.
Thus, in line with the work in \cite{suhtse:ICwithfeedback}, we conclude that when $\snr \geq \max \left \{ \Cc, \inr\right \}$ unilateral cooperation is too weak to allow for a full utilization of the channel resources, i.e., it leaves some system resources underutilized. 
{
In particular: 
\begin{itemize}

\item {\bf{Strong interference}} (i.e., $\inr > \snr$): 
in this regime neither the capacity region of the non-cooperative Gaussian IC \cite{etw} nor the capacity region of the non-causal Gaussian CIC \cite{riniJ1}, have bounds of the type  $2\Rp+\Rc$ and $\Rp+2\Rc$. 
It turns out that these types of bounds are not necessary for the GCCIC either. 

\item {\bf{Weak interference and strong cooperation}} (i.e., $\inr \leq \snr < \Cc$): 
for this regime the outer bound in \eqref{eq:OBsymmetricGaussian} equals (to within a constant gap) the outer bound on the capacity region for the non-causal Gaussian CIC \cite[Theorem III.1]{riniJ1}, which does not have bounds of the type $2\Rp+\Rc$ and $\Rp+2\Rc$ (see discussion in item~\ref{rem:samePerfasIdeal} in Section~\ref{subsect:OBGaussian}). 
In other words, in this regime the ideal non-causal cognition assumption at the CTx just provides a bounded rate increase compared to the more practical case of causal learning for the CTx through a noisy link.
It hence follows that for this regime unilateral cooperation allows to fully utilize the channel resources \cite{suhtse:ICwithfeedback}, i.e., the bounds of the type $2\Rp+\Rc$ and $\Rp+2\Rc$ are not active.
\item {\bf{Weak interference and weak cooperation}} (i.e., $\snr \geq \max \left \{ \Cc, \inr\right \}$): for this regime the capacity region of the non-cooperative Gaussian IC has bounds of the type $2\Rp+\Rc$ and $\Rp+2\Rc$ \cite{etw}, while the one of the non-causal Gaussian CIC does not \cite{riniJ1}. From our constant gap result in this region, it follows that $2\Rp+\Rc$ in \eqref{eq:OBsymmetricGaussian7} and $\Rp+2\Rc$ in \eqref{eq:OBsymmetricGaussian8} are both active. In other words, in this regime unilateral cooperation does not allow enough coordination among the sources which results in some `resource holes' as in the non-cooperative Gaussian IC.
\end{itemize}
}



%
%

\section{Conclusions}
\label{sec:Conclusion}
In this work we studied the two-user CCIC, an interference channel where one capable full-duplex source, i.e., the {\it cognitive} CTx, cooperates with / assists the other source, i.e., the {\it primary} PTx, to convey information.
In contrast to the original overlay cognitive paradigm, where the CTx a priori knows the message of the PTx, in the CCIC the CTx causally learns the primary's data through a noisy in-band link. 
We first derived two novel outer bounds of the type $2\Rp + \Rc$ and $\Rp + 2\Rc$ on the capacity region of the injective semi-deterministic channel with independent noises at the two pairs. We then designed a transmission strategy based on binning and superposition encoding, partial-decode-and-forward relaying and simultaneous decoding and we derived its achievable rate region. We finally evaluated the outer and lower bounds on the capacity for the practically relevant Gaussian noise case and we proved that our bounds are a constant number of bits apart from one another for the symmetric case (i.e., the two direct links and the two interfering links are of the same strength) in weak interference when the cooperation link is weaker than a given threshold. 
We showed that the two novel outer bounds of the type $2\Rp + \Rc$ and $\Rp + 2\Rc$ are active in weak interference when the cooperation link is weaker than the direct link, i.e., in this regime unilateral cooperation is too weak to allow for a full utilization of the channel resources.

\appendices

\section{Proof of the Markov chains in~\eqref{eq:MC with Wp} and~\eqref{eq:MC with Wc}} 
\label{app:FDG}

\begin{figure}
\centering
\includegraphics[width=0.8\textwidth]{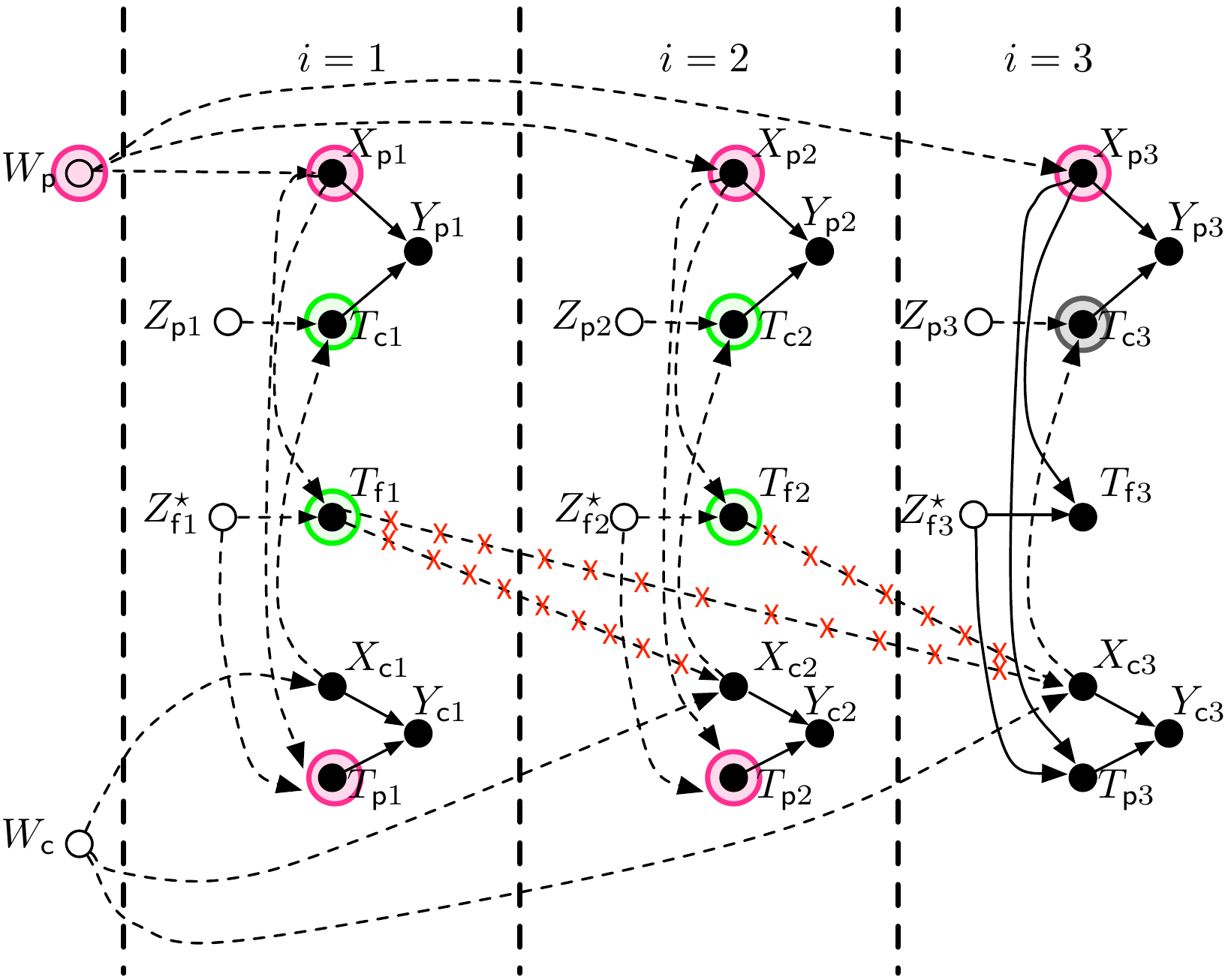}
\caption{Proof of the Markov chain in \eqref{eq:MC with Wp} using the FDG.}
\label{fig:MC1}
\end{figure}

\begin{figure}
\centering
\includegraphics[width=0.8\textwidth]{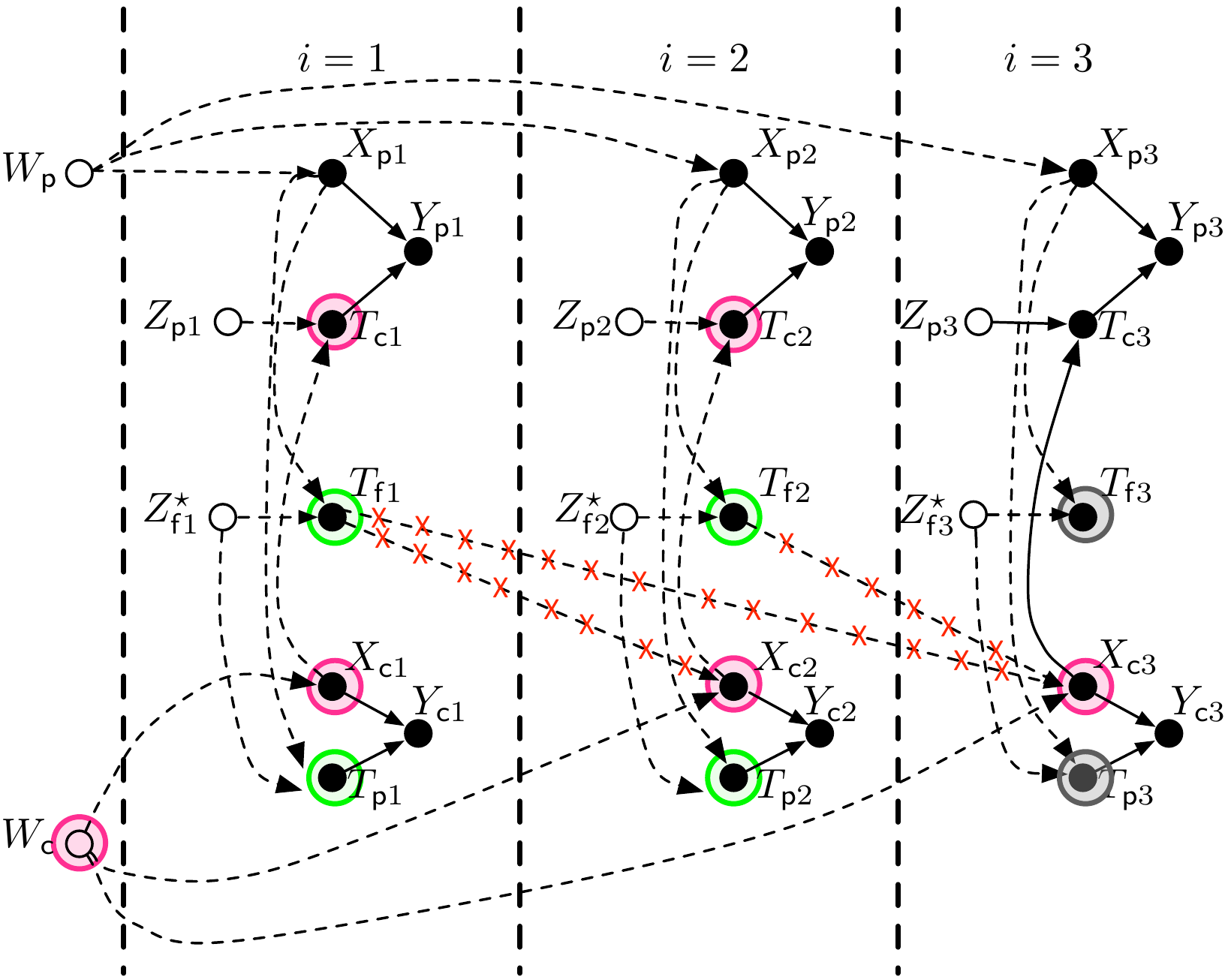}
\caption{Proof of the Markov chain in \eqref{eq:MC with Wc} using the FDG.}
\label{fig:MC2}
\end{figure}

We start by proving the two Markov chains in \eqref{eq:MC with Wp}-\eqref{eq:MC with Wc} by using the FDG \cite{KramerPhD}. Fig.~\ref{fig:MC1} proves the Markov chain in \eqref{eq:MC with Wp}, while Fig.~\ref{fig:MC2} the one in \eqref{eq:MC with Wc}. The two proofs, without loss of generality, consider the time instant $i=3$. According to \cite{KramerPhD}, we proceed through the following steps.
\begin{enumerate}
\item Draw the directed graph $\mathcal{G}_1$, which takes into consideration the dependence between the different random variables involved in the ISD CCIC considered. In particular, we define
\begin{align*}
Z^{\star}_{\mathsf{f}i} = \begin{bmatrix} \Zf_{i} \\ \Zc_{i} \end{bmatrix}
\end{align*}
to consider the fact that the noises at the CTx and at the CRx can be arbitrarily correlated
and we have
\begin{align*}
& \Xp_{i}=f(\Wp), \ \Yp_{i}=f(\Xp_{i},\Tc_{i}), \ \Tc_{i}= f (\Xc_{i},\Zp_{i}),\ \Yf_{i}=f(\Xp_{i},Z^{\star}_{\mathsf{f}i}),
\\ & \Xc_{i} = f(\Wc, \Yf^{i-1}), \ \Yc_{i}= f(\Xc_{i},\Tp_{i}), \ \Tp_{i}=f(\Xp_{i},Z^{\star}_{\mathsf{f}i}),
\end{align*}

where with $f$ we indicate that the left-hand side of the equality is a function of the random variables into the bracket.
\item In $\mathcal{G}_1$, highlight all the different nodes / random variables involved in the two Markov chains in \eqref{eq:MC with Wp}-\eqref{eq:MC with Wc} we aim to prove. In particular, the random variables circled in magenta, given those circled in green, should be proved to be independent of those circled in grey.
\item From the graph $\mathcal{G}_1$, consider the subgraph $\mathcal{G}_2$ which contains those edges and vertices encountered when moving backwards one or more edges starting from the colored (magenta, green and grey) random variables. The edges of the subgraph $\mathcal{G}_2$ are depicted with dashed black lines in Fig.~\ref{fig:MC1} and Fig.~\ref{fig:MC2} and the vertices in $\mathcal{G}_2$ are all those touched by a dashed black line.
\item From the graph $\mathcal{G}_2$, remove all the edges coming out from the random variables in green (those which are supposed to $d$-separate the random variables colored in magenta and grey). In Fig.~\ref{fig:MC1} and Fig.~\ref{fig:MC2}, this step is highlighted with red crosses on the edges which are removed. We let $\mathcal{G}_3$ be the subgraph obtained from $\mathcal{G}_2$ by removing all the edges with red crosses.
\item From $\mathcal{G}_3$, remove all the arrows on the edges, and obtain the undirected subgraph $\mathcal{G}_4$. In $\mathcal{G}_4$ it is easy to see that, by starting from any grey node, it is not possible to reach any magenta node. This concludes the proof of the two Markov chains in \eqref{eq:MC with Wp}-\eqref{eq:MC with Wc}.
\end{enumerate}

\section{Proof of the  sum-rate outer bound in~\eqref{eq:pv}}
\label{app:eq:pv}

By using the two Markov chains in \eqref{eq:MC with Wp}-\eqref{eq:MC with Wc} we can now derive the sum-rate outer bound
in~\eqref{eq:pv}. This bound was originally derived in \cite{PVIT11} for the case of independent noises; here we extend it to the case when only the noises at the different source-destination pairs are independent, i.e., $\mathbb{P}_{{Y}_{\mathsf{Fc}},\Yc|\Xp,\Xc}$ in \eqref{eq:distr} is not a product distribution.
By using Fano's inequality and by providing the same genie side information as in~\cite{PVIT11}, we have
\begin{align*}
& N(\Rp+\Rc-2\epsilon_N)
\\  &\leq
   I \left( \Wp; \Yp^N \right)
 + I \left( \Wc; \Yc^N \right) 
\\  &\leq
   I \left(  \Wp; \Yp^N, \Tp^N, \Yf^N \right)
  +I \left(  \Wc; \Yc^N, \Tc^N, \Yf^N \right) 
\\ = &
    H \left(  \Yp^N, \Tp^N, \Yf^N \right) - H \left(  \Yc^N, \Tc^N, \Yf^N|\Wc \right)
\\&+H \left(  \Yc^N, \Tc^N, \Yf^N \right) - H \left(  \Yp^N, \Tp^N, \Yf^N|\Wp \right).
\end{align*}
We now analyze and bound the two pairs of terms. First pair:
\begin{align*}
  &H \left(  \Yp^N, \Tp^N, \Yf^N \right) - H \left(  \Yc^N, \Tc^N, \Yf^N|\Wc \right)
\\ \stackrel{({\rm a})}{=}  &  \sum_{i\in[1:N]}
   H \left(  \Yp_{i}, \Tp_{i}, \Yf_{i}|\Yp^{i-1}, \Tp^{i-1}, \Yf^{i-1}  \right)
 - H \left(  \Yc_{i}, \Tc_{i}, \Yf_{i}|\Yc^{i-1}, \Tc^{i-1}, \Yf^{i-1}, \Wc,  \Xc^{i} \right)
\\ \stackrel{({\rm b})}{=} & \sum_{i\in[1:N]}
   H \left(  \Yp_{i}, \Tp_{i}, \Yf_{i}|\Yp^{i-1}, \Tp^{i-1}, \Yf^{i-1}  \right)
 - H \left(  \Tp_{i}, \Tc_{i}, \Yf_{i}|\Tp^{i-1}, \Tc^{i-1}, \Yf^{i-1}, \Wc,  \Xc^{i} \right)
\\ \stackrel{({\rm c})}{\leq}  & \sum_{i\in[1:N]}
\underbrace{
   H \left(  \Tp_{i}, \Yf_{i}|\Tp^{i-1}, \Yf^{i-1} \right)
 - H \left(  \Tp_{i}, \Yf_{i}|\Tp^{i-1}, \Yf^{i-1}, \Wc, \Tc^{i-1},\Xc^{i} \right)
 }_{\text{$=0$ because of~\eqref{eq:MC with Wc}}}
\\&+  \sum_{i\in[1:N]}
   H \left(  \Yp_{i}|\Tp_{i}, \Yf_{i}  \right)
 - H \left(  \Tc_{i}|\Tp^{i}, \Yf^{i}, \Wc, \Tc^{i-1},\Xc^{i},\Xp^{i} \right)
\\ \stackrel{({\rm d})}{=}  & \sum_{i\in[1:N]}
   H \left(  \Yp_{i}|\Tp_{i}, \Yf_{i}  \right)
 - H \left(  \Yp_{i}|\Tp_{i}, \Yf_{i},\Xp_{i},\Xc_{i} \right),
\end{align*}
where: the equality in $({\rm a})$ follows by applying the chain rule of the entropy and since, for the ISD CCIC, the encoding function $\Xc_{i}(\Wc,{Y}_{\mathsf{Fc}}^{i-1})$ is equivalent to $\Xc_{i}(\Wc,\Yf^{i-1})$; the equality in $({\rm b})$ is due to the fact that $\Yc$ is a deterministic function of $(\Xc,\Tp )$, which is invertible given $\Xc$; the inequality in $({\rm c})$ is due to the conditioning reduces entropy principle; the equality in $({\rm d})$ follows because of the ISD property of the channel and since the channel is memoryless. Second pair:
\begin{align*}
  &H \left(  \Yc^N, \Tc^N, \Yf^N \right) - H \left(  \Yp^N, \Tp^N, \Yf^N|\Wp \right)
\\ \stackrel{({\rm e})}{=}  &  \sum_{i\in[1:N]}
   H \left(  \Yc_{i}, \Tc_{i}, \Yf_{i}|\Yc^{i-1}, \Tc^{i-1}, \Yf^{i-1}  \right)
 - H \left(  \Yp_{i}, \Tp_{i}, \Yf_{i}|\Yp^{i-1}, \Tp^{i-1}, \Yf^{i-1}, \Wp,  \Xp^{i} \right)
\\ \stackrel{({\rm f})}{=} & \sum_{i\in[1:N]}
   H \left(  \Yc_{i}, \Tc_{i}, \Yf_{i}|\Yc^{i-1}, \Tc^{i-1}, \Yf^{i-1}  \right)
 - H \left(  \Tc_{i}, \Tp_{i}, \Yf_{i}|\Tc^{i-1}, \Tp^{i-1}, \Yf^{i-1}, \Wp,  \Xp^{i} \right)
\\ \stackrel{({\rm g})}{\leq}  & \sum_{i\in[1:N]}
\underbrace{
   H \left(  \Tc_{i}|\Tc^{i-1}, \Yf^{i-1} \right)
 - H \left(  \Tc_{i}|\Tc^{i-1}, \Yf^{i-1}, \Wp, \Tp^{i-1},\Xp^{i} \right)
 }_{\text{$=0$ because of~\eqref{eq:MC with Wp}}}
\\& + \sum_{i\in[1:N]}  H \left(  \Yf_{i} |\Tc_{i} \right) - H \left(\Yf_{i} |\Tc^{i}, \Yf^{i-1},\Wp, \Tp^{i-1},\Xp^{i} ,\Xc^{i}\right)
\\&+  \sum_{i\in[1:N]}
   H \left(  \Yc_{i}|\Tc_{i}, \Yf_{i}  \right)
 - H \left(  \Tp_{i}|\Tc^{i}, \Yf^{i}, \Wp, \Tp^{i-1},\Xp^{i},\Xc^{i} \right)
\\ \stackrel{({\rm h})}{=}  & \sum_{i\in[1:N]} H \left( \Yf_{i}|\Tc_{i}\right) - H \left( \Yf_{i} |\Tc_{i},\Xp_{i},\Xc_{i}\right) 
\\& + \sum_{i\in[1:N]}
   H \left(  \Yc_{i}|\Tc_{i}, \Yf_{i}  \right)
 - H \left(  \Yc_{i}|\Tc_{i}, \Yf_{i},\Xp_{i},\Xc_{i} \right),
\end{align*}
where: the equality in $({\rm e})$ follows by applying the chain rule of the entropy and since, given $\Wp$, $\Xp$ is uniquely determined; the equality in $({\rm f})$ is due to the fact that $\Yp$ is a deterministic function of $(\Xp,\Tc )$, which is invertible given $\Xp$; the inequality in $({\rm g})$ is due to the conditioning reduces entropy principle; the equality in $({\rm h})$ follows because of the ISD property of the channel and since the channel is memoryless.

By combining all the terms together, by introducing the time sharing random variable uniformly distributed over $[1:N]$ and independent of everything else, by dividing both sides by $N$ and taking the limit for $N\to\infty$ we get the bound in~\eqref{eq:pv}. We finally notice that by dropping the time sharing we do not decrease the bound.

\section{Proof of the outer bound in~\eqref{eq:boundRp+2Rc}}
\label{app:proofboundRp+2Rc}
By Fano's inequality, by considering that the messages $\Wp$ and $\Wc$ are independent and by giving side information similarly to \cite{PVIT11}, we have
\begin{align*}
&N(\Rp+2\Rc-3\epsilon_N)
 \\   &\leq 
  I \left( \Wp; \Yp^N \right)+     2I \left( \Wc; \Yc^N \right) 
\\  &\leq
        I \left( \Wp; \Yp^N, \Tp^N, \Yf^N \right) +   I \left( \Wc; \Yc^N \right)
    + I \left( \Wc; \Yc^N, \Tc^N, \Yf^N |\Wp\right) 
\\  &\leq
     H \left( \Yc^N \right )                   - H \left( \Yc^N, \Tc^N, \Yf^N |\Wp,\Wc \right)
\\&+ H \left( \Yc^N, \Tc^N, \Yf^N |\Wp \right) - H \left( \Yp^N, \Tp^N, \Yf^N |\Wp \right)
\\&+ H \left( \Yp^N, \Tp^N, \Yf^N \right)      - H \left( \Yc^N| \Wc \right ).
\end{align*}

We now analyze each pair of terms. In particular, we proceed similarly as we did to prove the outer bound $2 \Rp+\Rc$ in \eqref{eq:bound2Rp+Rc}.
\\
First pair:
\begin{align*}
  &H \left( \Yc^N \right )- H \left( \Yc^N, \Tc^N, \Yf^N |\Wp,\Wc \right)
\\ {=} & \sum_{i\in[1:N]} 
    H \left(\Yc_i |\Yc^{i-1}\right)
 - H \left(\Yc_{i}, \Tc_{i}, \Yf_{i} | \Wp,\Wc,\Yc^{i-1},\Tc^{i-1},\Yf^{i-1}, \Xp^{i}, \Xc^{i}\right)
\\ {\leq} & \sum_{i\in[1:N]} 
    H \left(\Yc_i \right)
 - H \left(\Yc_{i}, \Tc_{i}, \Yf_{i} | \Wp,\Wc,\Yc^{i-1},\Tc^{i-1},\Yf^{i-1}, \Xp^{i}, \Xc^{i}\right)
\\ {=} & \sum_{i\in[1:N]} 
    H \left(\Yc_i \right)
  - H \left(\Yc_{i}, \Tc_{i}, \Yf_{i} | \Xp_{i}, \Xc_{i}\right).
\end{align*}
Second pair:
\begin{align*}
  &H \left( \Yc^N, \Tc^N, \Yf^N |\Wp \right)
 - H \left( \Yp^N, \Tp^N, \Yf^N|\Wp \right)
\\ {=} & \sum_{i\in[1:N]}
   H \left( \Yc_{i}, \Tc_{i}, \Yf_{i}|\Yc^{i-1}, \Tc^{i-1},  \Yf^{i-1}, \Wp , \Xp^{i}\right)  -   H \left( \Yp_{i}, \Tp_{i}, \Yf_{i}|\Yp^{i-1}, \Tp^{i-1},  \Yf^{i-1}, \Wp , \Xp^{i}\right)
\\ {=} & \sum_{i\in[1:N]}
   H \left( \Yc_{i}, \Tc_{i}, \Yf_{i}|\Yc^{i-1}, \Tc^{i-1},  \Yf^{i-1}, \Wp , \Xp^{i}\right)  -    H \left( \Tc_{i}, \Tp_{i}, \Yf_{i}|\Tc^{i-1}, \Tp^{i-1},  \Yf^{i-1}, \Wp , \Xp^{i}\right)
\\ {\leq} & \sum_{i\in[1:N]}
   \underbrace{ H \left( \Tc_{i}|           \Tc^{i-1},  \Yf^{i-1}\right)
 - H \left( \Tc_{i} |\Tc^{i-1}, \Tp^{i-1}, \Yf^{i-1}, \Wp, \Xp^{i} \right) }_{\text{$= 0$ because of~\eqref{eq:MC with Wp}}}
\\& + \sum_{i\in[1:N]}
    H \left( \Yf_{i} |\Tc_{i}, \Xp_{i} \right) 
  - H \left( \Yf_{i} |\Tc^{i}, \Tp^{i-1}, \Yf^{i-1}, \Wp, \Xp^{i}\right)
\\ &+ \sum_{i\in[1:N]}
   H \left( \Yc_i|\Tc_{i},  \Yf_{i}, \Xp_{i}\right)
-    H \left( \Tp_i|\Tc^{i}, \Tp^{i-1},  \Yf^{i}, \Wp , \Xp^{i}, \Xc^{i}\right)
\\ {=} & 
\sum_{i\in[1:N]}
    H \left( \Yf_{i} |\Tc_{i}, \Xp_{i} \right) 
  - H \left( \Yf_{i} |\Tc_{i}, \Xp_{i} \right)
\\& + \sum_{i\in[1:N]}
   H \left( \Yc_i|\Tc_{i},  \Yf_{i}, \Xp_{i}\right)
-  H \left( \Yc_i|\Tc_{i},  \Yf_{i}, \Xc_{i}, \Xp_i\right)
\\ = & 
 \sum_{i\in[1:N]}
   H \left( \Yc_i|\Tc_{i},  \Yf_{i}, \Xp_{i}\right)
-  H \left( \Yc_i|\Tc_{i},  \Yf_{i}, \Xc_{i}, \Xp_i\right).
\end{align*}
Third pair: since
\begin{align*}
&H \left ( \Yc^N|\Wc \right)
\\ =&    \sum_{i\in[1:N]} H \left ( \Yc_{i}|\Yc^{i-1},\Wc\right)
\\ \geq & \sum_{i\in[1:N]} H \left ( \Yc_{i}|\Yc^{i-1},\Wc,\Yf^{i-1},\Xc^{i}\right) 
\\ = &    \sum_{i\in[1:N]} H \left ( \Tp_i|\Tp^{i-1},\Wc,\Yf^{i-1},\Xc^{i}\right) 
\\ = &    \sum_{i\in[1:N]} H \left ( \Tp_i|\Tp^{i-1},\Yf^{i-1}\right),
\end{align*}
where the last equality follows because of~\eqref{eq:MC with Wc},
then
\begin{align*}
  &H \left( \Yp^N, \Tp^N, \Yf^N \right)      - H \left( \Yc^N| \Wc \right )
\\  &\leq \sum_{i\in[1:N]}
   H \left( \Yp_{i}, \Tp_{i}, \Yf_{i} | \Yp^{i-1}, \Tp^{i-1}, \Yf^{i-1}\right)      - H \left ( \Tp_i|\Tp^{i-1},\Yf^{i-1}\right)
\\  &\leq \sum_{i\in[1:N]}
   H \left( \Yp_{i},  \Yf_{i} | \Tp_{i}\right).
\end{align*}

By combining everything together, by introducing the time sharing random variable uniformly distributed over $[1:N]$ and independent of everything else, by dividing both sides by $N$ and taking the limit for $N\to\infty$ we get the bound in \eqref{eq:boundRp+2Rc}. We finally notice that by dropping the time sharing we do not decrease the bound.

\section{Evaluation of the outer bounds in \eqref{eq:outknownbefore}, \eqref{eq:bound2Rp+Rc} and \eqref{eq:boundRp+2Rc} for the GCCIC}
\label{app:OBGaussianGeneral}
By defining $\mathbb{E} \left [ \Xp \Xc^* \right ] := \rho: |\rho| \in [0,1]$ we obtain: from the cut-set bounds in~\eqref{eq:cutset1 a}-\eqref{eq:cutset2}
\begin{subequations}
\label{eq:knownOBgeneralGaussian}
\begin{align}
 \Rp    &\leq  \log \left( 1 + \left( \Cc + \Sp \right) \left( 1-|\rho|^2\right)\right) 
\notag\\&\stackrel{|\rho|=0}{\leq} \log \left( 1+\Cc+\Sp \right),
\\ \Rp  &\leq \log \left( 1+ \Sp + \Ic + 2 \sqrt{\Sp \Ic } \mathfrak{R} \left \{ \rho {\rm e}^{-{\rm j}\theta_{\mathsf{c}}} \right \} \right)  
\notag\\&\stackrel{\rho=\eac}{\leq}  \log\left ( 1+\left (\sqrt{\Sp} +  \sqrt{\Ic} \right)^2\right ),
\\ \Rc  &\leq \log \left( 1+ \left( 1- |\rho|^2 \right) \Sc \right)
\notag\\&\stackrel{|\rho|=0}{\leq}  \log \left( 1 +  \Sc \right).
\end{align}
From the bounds in~\eqref{eq:tuni1}-\eqref{eq:tuni2} we get
\begin{align}
\Rp + \Rc &\leq  
   \log \left( 1 + \frac{ \left(\Sp +\Cc \right) \left( 1-|\rho|^2\right)}{1+ \Ip\left( 1-|\rho|^2\right)}  \right)
 + \log \left( 1+ \Sc + \Ip + 2 \sqrt{\Sc \Ip } \mathfrak{R} \left \{ \rho \eap \right \} \right)  \nonumber
\\& \stackrel{({\rm a})}{\leq}  
   \log  \left( 1 + \frac{ \Sp +\Cc}{1+ \Ip}  \right) 
+  \log  \left( 1+ \left (\sqrt{\Sc} + \sqrt{\Ip} \right)^2  \right)
\\& =
\underbrace{   
   \log  \left( 1 + \frac{ \Sp}{1+ \Ip}  \right) 
+  \log  \left( 1+ \left (\sqrt{\Sc} + \sqrt{\Ip} \right)^2  \right)}_{\text{as no cooperation / $\Cc=0$}}
\notag
\\& 
+ \underbrace{  \log  \left(1+ \frac{\Cc}{1+ \Ip+ \Sp}  \right)}_{\text{increasing in $\Cc$}},
\notag
\end{align}
and
\begin{align}
\Rp + \Rc  & \leq \log  \left( 1 + \frac{ \Sc \left( 1-|\rho|^2\right)}{1+ \Ic \left( 1-|\rho|^2\right)}  \right) + \log \left( 1+ \Sp + \Ic + 2 \sqrt{\Sp \Ic } \mathfrak{R} \left \{ \rho {\rm e}^{-{\rm j}\theta_{\mathsf{c}}} \right \} \right) \nonumber
\\& \stackrel{({\rm b})}{\leq}
\underbrace{ 
 \log  \left( 1 + \frac{ \Sc}{1+ \Ic}  \right) + \log  \left( 1+ \left(\sqrt{\Sp} + \sqrt{\Ic} \right)^2  \right)}_{\text{as no cooperation / $\Cc=0$}},
\end{align}
where the inequality in $({\rm{a}})$ follows by evaluating the first logarithm in $|\rho|=0$ and the second logarithm in $\rho={\rm e}^{-{\rm j}\theta_{\mathsf{p}}}$ and the inequality in $({\rm{b}})$ follows by evaluating the first logarithm in $|\rho|=0$ and the second logarithm in $\rho=\eac$.
Finally, from the bound in~\eqref{eq:pv} we obtain
\begin{align}
 \Rp + \Rc & \leq \log \left( 1 + \frac{\Sp+\Ic+ 2 \sqrt{\Sp \Ic } \mathfrak{R} \left \{ \rho {\rm e}^{-{\rm j}\theta_{\mathsf{c}}} \right \}+\left( 1-|\rho|^2\right) \left( \Ip \Ic + \Cc \Ic \right)}{1+\Cc+\Ip} \right) \nonumber
\\& + \log \left( 1 + \frac{\Sc+\Ip+ 2 \sqrt{\Sc \Ip } \mathfrak{R} \left \{ \rho \eap \right \}+\left( 1-|\rho|^2\right) \left( \Ip \Ic + \Cc \Sc \right)}{1+\Cc+\Ic + \Ic \Cc \left( 1-|\rho|^2\right)} \right) \nonumber
\\& + \log \left( 1 + \frac{\Cc + \Cc \Ic \left( 1-|\rho|^2\right)}{1+\Ic} \right) \nonumber
\\ & = \log \left( 1 + \frac{\Sp+\Ic+ 2 \sqrt{\Sp \Ic } \mathfrak{R} \left \{ \rho {\rm e}^{-{\rm j}\theta_{\mathsf{c}}} \right \}+\left( 1-|\rho|^2\right) \left( \Ip \Ic + \Cc \Ic \right)}{1+\Cc+\Ip} \right) \nonumber
\\& + \log \left( 1 + \frac{\Cc+\Sc+\Ip+ 2 \sqrt{\Sc \Ip } \mathfrak{R} \left \{ \rho \eap \right \}+\left( 1-|\rho|^2\right) \left( \Ip \Ic + \Cc \Sc + \Ic \Cc\right)}{1+\Ic } \right) \nonumber
\\ &  \stackrel{({\rm c})}{\leq} \log \left( 1 + \Ic + \frac{\Sp}{1+\Cc+\Ip} \right)+ \log \left( 1 + \Cc + \Ip+ \frac{\Sc\left( 1+\Cc\right) }{1+\Ic} \right) +2 \log(2),
\\ & =
\underbrace{ \log \left(  1+\Cc+\Ip+ \frac{\Sp}{1 + \Ic} \right)+ \log \left( 1+\Ic+ \Sc\frac{1+\Cc}{1 + \Cc + \Ip} \right)
}_{\text{increasing in $\Cc$}}
  +2 \log(2),
\notag
\end{align}
where the  inequality in $({\rm{c}})$ follows by: (i) evaluating the first term of the first logarithm in $\rho=\eac$ and the second term of the first logarithm in $|\rho|=0$; (ii) evaluating the first term of the second logarithm in $\rho={\rm e}^{-{\rm j}\theta_{\mathsf{p}}}$ and the second term of the second logarithm in $|\rho|=0$; (iii) since $\log \left (1+\left( \sqrt{|a|^2} + \sqrt{|b|^2}\right)^2 \right ) \leq \log \left( 1+ |a|^2 + |b|^2\right) + \log(2)$.

We now evaluate the new outer bounds in Theorem~\ref{th:main th 2r1+r2 and r1+2r2} and we get
\begin{align}
 2 \Rp + \Rc& \leq \log \left( 1+ \Sp + \Ic + 2 \sqrt{\Sp \Ic } \mathfrak{R} \left \{ \rho {\rm e}^{-{\rm j}\theta_{\mathsf{c}}} \right \} \right) + \log \left( 1 + \frac{\Sp \left( 1-|\rho|^2\right)}{1 + \left( \Cc+\Ip\right) \left( 1-|\rho|^2\right)} \right) \nonumber
\\& + \log \left( 1 + \frac{\Cc+\Sc+\Ip+ 2 \sqrt{\Sc \Ip } \mathfrak{R} \left \{ \rho \eap \right \}+\left( 1-|\rho|^2\right) \left( \Ip \Ic + \Cc \Sc + \Ic \Cc\right)}{1+\Ic } \right) \nonumber
\\ & \stackrel{({\rm{d}})}{\leq}
 \log\left ( 1+\left (\sqrt{\Sp} +  \sqrt{\Ic} \right)^2\right ) + \log \left( 1 + \frac{\Sp}{1+\Ip+\Cc} \right) 
\nonumber\\&
 + \log \left( 1 + \Cc + \Ip+ \frac{\Sc\left( 1+\Cc\right) }{1+\Ic} \right) + \log(2),
\label{eq:2R1 + R2 Gaussian}
\\ &  =
\underbrace{
   \log \left( 1+\frac{\Sc}{1+\Ic}\right)
  +\log\left ( 1+\left (\sqrt{\Sp} +  \sqrt{\Ic} \right)^2\right )}_{\text{as no cooperation / $\Cc=0$}}
\notag
\\&
 +\underbrace{
   \log \left( \frac{1+\Ip+ \Sp}{1+\Ic+\Sc} \right) 
 + \log \left( 1+\frac{\Cc}{1+\Ip+ \Sp}\right)
 + \log \left( 1+\Ic+ \Sc\frac{1+\Cc}{1+\Ip+\Cc} \right)}_{\text{increasing in $\Cc$}}
 + \log(2),
\notag
 \end{align}
where the inequality in $({\rm{d}})$ follows by (i) evaluating the first logarithm in $\rho=\eac$, (ii) evaluating the second logarithm in $|\rho|=0$, (iii) evaluating the first term of the third logarithm in $\rho={\rm e}^{-{\rm j}\theta_{\mathsf{p}}}$ and the second term of the third logarithm in $|\rho|=0$ and (iv) since $\log \left (1+\left( \sqrt{|a|^2} + \sqrt{|b|^2}\right)^2 \right ) \leq \log \left( 1+ |a|^2 + |b|^2\right) + \log(2)$.
Similarly,
\begin{align}
 \Rp+2\Rc & \leq \log \left( 1+ \Sc + \Ip + 2 \sqrt{\Sc \Ip } \mathfrak{R} \left \{ \rho \eap \right \} \right) + \log \left( 1 + \frac{ \Sc \left( 1-|\rho|^2\right)}{1+ \Ic\left( 1-|\rho|^2\right)}  \right) \nonumber
\\& + \log \left( 1 + \frac{\Sp+\Ic+ 2 \sqrt{\Sp \Ic } \mathfrak{R} \left \{ \rho {\rm e}^{-{\rm j}\theta_{\mathsf{c}}} \right \}+\left( 1-|\rho|^2\right) \left( \Ip \Ic + \Cc \Ic \right)}{1+\Cc+\Ip} \right) \nonumber
\\& + \log \left( 1+\frac{\Cc}{1+\Ip}\right) \nonumber
\\ & \stackrel{({\rm{e}})}{\leq} \log  \left( 1+ \left (\sqrt{\Sc} + \sqrt{\Ip} \right)^2  \right)+ \log \left( 1 + \frac{ \Sc}{1+ \Ic}  \right) \nonumber
\\ & +\log \left( 1 + \Ic + \frac{\Sp}{1+\Cc+\Ip} \right)+ \log \left( 1+\frac{\Cc}{1+\Ip}\right)+\log(2),
\label{eq:R1 + 2R2 Gaussian}
\\ &=\underbrace{ 
   \log \left( 1+\frac{\Sp}{1+\Ip}\right)
 + \log  \left( 1+ \left (\sqrt{\Sc} + \sqrt{\Ip} \right)^2  \right)}_{\text{as no cooperation / $\Cc=0$}}
\notag
\\&
+\underbrace{
  \log \left( \frac{1+ \Ic + \Sc}{1+\Ip+\Sp} \right) 
+ \log \left( 1+\Cc+\Ip + \frac{\Sp}{1 + \Ic} \right)}_{\text{increasing in $\Cc$}}
+\log(2),
\notag
\end{align}
\end{subequations}
where the inequality in $({\rm{e}})$ follows by: (i) evaluating the first logarithm in $\rho={\rm e}^{-{\rm j}\theta_{\mathsf{p}}}$, (ii) evaluating the second logarithm in $|\rho|=0$ and (iii) evaluating the first term of the third logarithm in $\rho=\eac$ and the second term of the third logarithm in $|\rho|=0$ and again (iv) since $\log \left (1+\left( \sqrt{|a|^2} + \sqrt{|b|^2}\right)^2 \right ) \leq \log \left( 1+ |a|^2 + |b|^2\right) + \log(2)$.


\section{Achievable Scheme Based on Superposition Coding and Binning}
\label{sec:allachschemsappDPC}

\begin{figure}[!h]
\centering
\includegraphics[width=\textwidth]{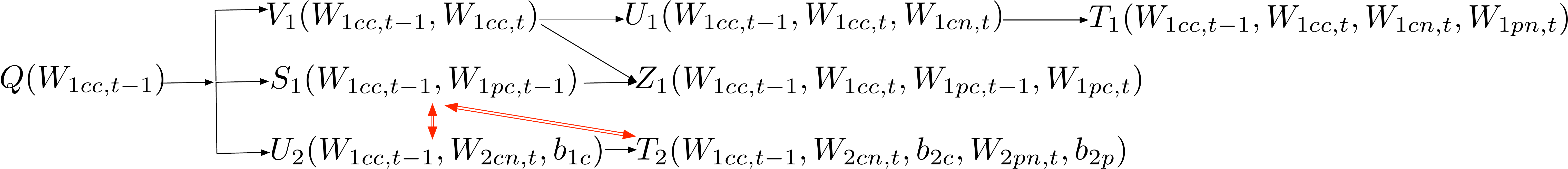}
\caption{Achievable scheme based on binning and superposition coding.}
\label{fig:achdpc}
\end{figure}

We specialize the `binning+superposition' achievable scheme in~\cite[Section V]{YANG-TUNINETTI}.
In~\cite[Thereom V.1]{YANG-TUNINETTI} the network comprises four nodes numbered from 1 to 4; nodes 1 and 2 are sources and nodes 3 and 4 are destinations; source node $j\in[1:2]$, with input to the channel $X_j$ and output from the channel $Y_j$, has a message $W_j$ for node $j+2$; destination node $j\in[3:4]$ has channel output $Y_{j}$ from which it decodes the message $W_{j-2}$. 
Both users use rate splitting, where the messages of user 1 / primary are both non-cooperative and cooperative, while the messages of user 2 / cognitive are non-cooperative.
In~\cite[Section V]{YANG-TUNINETTI}, we set $Y_1=S_2=V_2=Z_2=\emptyset$, 
i.e., then $R_1=R_{11c}+R_{10c}+R_{10n}+R_{11n},  \ R_2=R_{22n}+R_{20n}$, to obtain a scheme that comprises:
a cooperative common message (carried by the pair $(Q,V_1)$ at rate $R_{10c}$) for user 1,
a cooperative private message (carried by the pair $(S_1,Z_1)$ at rate $R_{11c}$) for user 1,
a non-cooperative common message (carried by $U_1$ at rate $R_{10n}$) for user 1,
a non-cooperative private message (carried by $T_1$ at rate $R_{11n}$) for user 1,
a non-cooperative common message  (carried by $U_2$ at rate $R_{20n}$) for user 2 and
a non-cooperative private message (carried by $T_2$ at rate $R_{22n}$) for user 2.
Here the pair $(Q,S_1)$ carries the `past cooperative messages', and the pair $(V_1,Z_1)$ the `new cooperative messages' in a block Markov encoding scheme.  
The channel inputs are functions of the auxiliary random variables, where $X_1$ is a function of $(Q,S_1,Z_1,V_1,U_1,T_1)$ and $X_2$ is a function of $(Q,S_1,U_2,T_2)$.

\paragraph{Input distributions}
The set of possible input distributions is
\begin{align}
 &P_{Q,S_1,V_1,Z_1,U_1,T_1,X_1, U_2,T_2,X_2} \nonumber
\\& = P_{Q} P_{V_1|Q} P_{U_1,T_1|Q,V_1}
  P_{S_1|Q}P_{Z_1|Q,S_1,V_1}
  P_{U_2,T_2|S_1,Q}
  P_{X_1|Q,S_1,Z_1,V_1,U_1,T_1}
  P_{X_2|Q,S_1,U_2,T_2}.
\label{eq:input pdf dpc}
\end{align}
A schematic representation of the achievable scheme is given in Fig.~\ref{fig:achdpc}, where a black arrow indicates superposition coding and a red arrow indicates binning. 
Note that the difference between Fig.~\ref{fig:achdpc} and \cite[Fig. 10]{ourITjournal} is that in the achievable scheme in  \cite[Fig. 10]{ourITjournal} we set $X_1=T_1$, $X_2=T_2$ and $U_1=S_1=Z_1=\emptyset$, i.e., user 1 / primary only used a common cooperative message and a private non-cooperative message, while here also private cooperative as well as common non-cooperative messages are used.
Similarly, the difference  between Fig.~\ref{fig:achdpc} and \cite[Fig. 11]{ourITjournal} is that in the achievable scheme in \cite[Fig. 11]{ourITjournal} we set $U_1=T_1=\emptyset$, i.e., the messages of user 1 / primary were only cooperative, while here we consider also non-cooperative (both private and common) messages for user 1 / primary.


\paragraph{Encoding}
The codebooks are generated as follows:
first the codebook $Q$ is generated;  
then the codebook $V_1$ is superposed to $Q$, after the codebook $U_1$ is superposed to $(Q,V_1)$ and finally  the codebook $T_1$ is superposed to $(Q,V_1,U_1)$;
independently of $(V_1,U_1,T_1)$, the codebook $S_1$ is superposed to $Q$ and then the codebook $Z_1$ is superposed to $(Q,S_1,V_1)$;
independently of $(S_1,Z_1,V_1,U_1,T_1)$, the codebook $U_2$ is superposed to $Q$ and then the codebook $T_2$ is superposed to $(Q,U_2)$.
With this random coding codebook generation, the pair $(U_2,T_2)$ is independent of $S_1$ conditioned on $Q$.
\cite[Theorem V.1]{YANG-TUNINETTI} involves several binning steps to allow for a large set of input distributions. Here the only binning steps are for $(U_2,T_2)$ against $S_1$.

\begin{subequations}
\label{eq:DPC unified}
We use a block Markov coding scheme to convey the message of user~1 to user~2.
In particular, at the end of any given time slot in a block Markov coding scheme, encoder~2 knows $(Q,S_1,U_2,T_2)$ and decodes $(V_1,Z_1)$ from its channel output; the decoded pair $(V_1,Z_1)$ becomes the pair $(Q,S_1)$ of the next time slot; then, at the beginning of each time slot, encoder~2, by binning, finds the new pair $(U_2,T_2)$ that is jointly typical with $(Q,S_1)$; for this to be possible, we must generate several $(U_2,T_2)$ sequences for each message of user 2 so as to be able to find one pair to send with the correct joint distribution with $(Q,S_1)$; 
this entails the rate penalties in \cite[eq(20)]{YANG-TUNINETTI} for user~1 and then again \cite[eq(20)]{YANG-TUNINETTI} for user~2 by swapping the role of the subscripts~1 and~2, with $S_2=Z_2=V_2=\emptyset$, i.e.,
\begin{align}
   R_{20n}^{'} + R_{22n}^{'} &\geq I(U_2,T_2;S_1|Q), \label{eq:const for dpc:node2 u2 t2}
\\ R_{20n}^{'}               &\geq I(U_2;S_1|Q).     \label{eq:const for dpc:node2 u2}   
\end{align}

\paragraph{Decoding}
The cooperative source uses the partial-decode-and-forward strategy and the destinations backward decoding.
There are three decoding nodes in the network and therefore three groups of rate constraints.
These are:
\begin{itemize}
\item
Node2 / CTx jointly decodes $(V_1,Z_1)$ from its channel output with knowledge of the indices in $(Q,S_1,U_2,T_2,X_2)$. 
Successful decoding is possible if (use \cite[eq(21)]{YANG-TUNINETTI} by swapping the role of the subscripts~1 and~2, with $S_2=Z_2=V_2=\emptyset$ and with $V_1$ independent of $S_1$)
\begin{align}
R_{10c}+R_{11c} & \leq I(Y_2;Z_1,V_1|U_2,T_2,X_2,S_1,Q), \label{eq:c1}
 \\           R_{11c} &\leq I(Y_2;Z_1|U_2,T_2,X_2,S_1,Q,V_1). \label{eq:c2}
\end{align}
\item
Node3 / PRx jointly decodes $(Q,S_1,U_2,U_1,T_1)$ from its channel output, with knowledge of some message indices in $(V_1,Z_1)$, by treating $T_2$ as noise. Successful decoding is possible if 
(see \cite[eq(22)]{YANG-TUNINETTI} with $S_2=Z_2=V_2=\emptyset$)
\begin{align}
&R_{10c} + R_{10n} + R_{11n} + R_{20n} + R_{11c}  \leq  I(Y_3;Q,V_1,U_1,T_1,S_1,Z_1,U_2) \nonumber
\\& \qquad -(R_{20n}^{'}-I(U_2;S_1|Q)),  \label{eq:c3}
\\ &R_{10n} + R_{11n} + R_{20n} + R_{11c} \leq I(Y_3;U_1,T_1,S_1,Z_1,U_2|Q,V_1)  \nonumber
\\& \qquad- (R_{20n}^{'} - I(U_2;S_1|Q)),    \label{eq:c4}
\\ &R_{10n} + R_{11n} + R_{11c} \leq I(Y_3;U_1,T_1,S_1,Z_1|Q,V_1,U_2),    \label{eq:c5}
\\ &R_{11n} + R_{20n} + R_{11c} \leq I(Y_3;T_1,S_1,Z_1,U_2|Q,V_1,U_1) -(R_{20n}^{'}-I(U_2;S_1|Q)),    \label{eq:c6}
\\ &R_{10n} + R_{11n} + R_{20n} \leq I(Y_3;U_1,T_1,U_2|Q,S_1,Z_1,V_1) -(R_{20n}^{'}-I(U_2;S_1|Q)),    \label{eq:c7}
\\   &        R_{11n} + R_{11c} \leq I(Y_3;T_1,S_1,Z_1|Q,V_1,U_1,U_2), \label{eq:c8}
\\  &         R_{20n} + R_{11c} \leq I(Y_3;      S_1,Z_1,U_2|Q,V_1,U_1,T_1)-(R_{20n}^{'}-I(U_2;S_1|Q)), \label{eq:c9}
\\     &      R_{10n} + R_{11n} \leq I(Y_3;      U_1,T_1|Q,S_1,Z_1,V_1,U_2),\label{eq:c10}
\\           &R_{11n} + R_{20n} \leq I(Y_3;      T_1,U_2|Q,S_1,Z_1,V_1,U_1)-(R_{20n}^{'}-I(U_2;S_1|Q)),\label{eq:c11}
\\                   &  R_{11c} \leq I(Y_3;      S_1,Z_1|Q,V_1,U_1,T_1,U_2),\label{eq:c12}
\\                  &   R_{11n} \leq I(Y_3;      T_1|Q,S_1,Z_1,V_1,U_1,U_2).\label{eq:c13}
\end{align}

\item
Node4 / CRx jointly decodes $(Q,U_1,U_2,T_2)$ from its channel output, with knowledge of some message index in $V_1$, by treating $Z_1$ and $T_1$ as noise (recall that the pair $(U_2,T_2)$ has been precoded/binned against $S_1$).  Successful decoding is possible if 
(see \cite[eq(22)]{YANG-TUNINETTI}, with the role of the users swapped, where only the bounds in \cite[eq(22a)]{YANG-TUNINETTI}, \cite[eq(22h)]{YANG-TUNINETTI}, \cite[eq(22i)]{YANG-TUNINETTI}, \cite[eq(22j)]{YANG-TUNINETTI}, and \cite[eq(22k)]{YANG-TUNINETTI} remain after setting several auxiliary random variables to zero and removing the redundant constraints)
\begin{align}
&   R_{10c}+R_{20n}+ R_{22n}+R_{10n} \leq I(Y_4;Q,U_2,T_2,V_1,U_1)-(R_{20n}^{'} + R_{22n}^{'}), \label{eq:c14}
\\ &   R_{20n}+ R_{22n}+R_{10n} \leq I(Y_4;U_2,T_2,U_1|Q,V_1)-(R_{20n}^{'} + R_{22n}^{'}), \label{eq:c15}
\\         &R_{20n}+ R_{22n} \leq I(Y_4;U_2,T_2|Q,V_1,U_1)-(R_{20n}^{'} + R_{22n}^{'}), \label{eq:c16}
\\         &R_{22n}+ R_{10n} \leq I(Y_4;T_2,U_1|Q,U_2,V_1)- R_{22n}^{'}, \label{eq:c17}
\\                  &R_{22n} \leq I(Y_4;T_2|Q,U_2,V_1,U_1)-R_{22n}^{'}.                 \label{eq:c18}       
\end{align}
\end{itemize}
\end{subequations}

\paragraph{Compact region}
Instead of applying the Fourier-Motzkin Elimination (FME) directly on the general achievable rate region, 
in the following we apply the FME on two particular cases, namely the case when $S_1=Z_1=\emptyset$, i.e., no private cooperative messages for the PTx and the case $U_1=\emptyset$, i.e., no common non-cooperative message for the PTx. The first case is analyzed in Section \ref{sub:noPCPTx}, while the second case is analyzed in Section \ref{sub:noCNCPTx}. For both cases we take the constraints in~\eqref{eq:const for dpc:node2 u2 t2} and~\eqref{eq:const for dpc:node2 u2} to hold with equality (i.e., $R_{20n}^{'} = I(U_2;S_1|Q), \ R_{22n}^{'} = I(S_1;T_2|Q,U_2)$).

\subsection{FME on the achievable rate region when $S_1=Z_1=\emptyset$} 
\label{sub:noPCPTx}
We set $S_1=Z_1=\emptyset$ in the achievable rate region. 
After FME of the achievable region in~\eqref{eq:DPC unified} with $S_1=Z_1=\emptyset$ (see also \cite[eq(8)]{YANG-TUNINETTI}), we get
\begin{subequations}
\begin{align}
  &R_1 \leq    {\rm eq\eqref{eq:c3}},
\label{eq:c1C1} 
\\&R_1 \leq     {\rm eq\eqref{eq:c1}} + {\rm eq\eqref{eq:c5}},
\label{eq:c2C1}
\\&R_2 \leq    {\rm eq\eqref{eq:c16}},
\label{eq:c3C1}
\\&R_1	+R_2	\leq	 {\rm eq\eqref{eq:c3}} + {\rm eq\eqref{eq:c18}} ,
\label{eq:c4C1}
\\&R_1	+R_2	\leq	{\rm eq\eqref{eq:c8}} +  {\rm eq\eqref{eq:c14}},
\label{eq:c5C1}
\\&R_1	+R_2	\leq		{\rm eq\eqref{eq:c1}} + {\rm eq\eqref{eq:c7}}+ {\rm eq\eqref{eq:c18}},
 \label{eq:c6C1}
\\&R_1	+R_2	\leq		{\rm eq\eqref{eq:c1}}+{\rm eq\eqref{eq:c8}}+{\rm eq\eqref{eq:c15}},
 \label{eq:c7C1}
\\&R_1	+R_2	\leq		{\rm eq\eqref{eq:c1}}+{\rm eq\eqref{eq:c11}}+{\rm eq\eqref{eq:c17}}, 
\label{eq:c8C1}
\\&2R_1 +R_2 \leq		{\rm eq\eqref{eq:c1}}+{\rm eq\eqref{eq:c8}} + {\rm eq\eqref{eq:c3}} +{\rm eq\eqref{eq:c17}},
 \label{eq:c9C1}
\\&2R_1	+R_2	\leq 	2 \cdot {\rm eq\eqref{eq:c1}}+{\rm eq\eqref{eq:c8}}+ {\rm eq\eqref{eq:c7}}+{\rm eq\eqref{eq:c17}},
 \label{eq:c10C1}
\\&R_1 +2R_2\leq		{\rm eq\eqref{eq:c11}} +{\rm eq\eqref{eq:c17}} +{\rm eq\eqref{eq:c14}}, 
\label{eq:c11C1}
\\&R_1 +2R_2\leq		{\rm eq\eqref{eq:c1}} +{\rm eq\eqref{eq:c11}}+ {\rm eq\eqref{eq:c18}} + {\rm eq\eqref{eq:c15}},
 \label{eq:c12C1}
\end{align}
\label{eq:noPCPTx}
\end{subequations}
for all distributions that factor as~\eqref{eq:input pdf dpc} and by setting $S_1=Z_1=\emptyset$ in all the mutual information terms.

We identify Node1 with the PTx (i.e., $\Xp=X_1$), Node2 with the CTx  (i.e., $\Xc=X_2, \Yf=Y_2$), Node3 with the PRx  (i.e., $\Yp=Y_3$) and Node4 with the CRx (i.e., $\Yc=Y_4$). 
For the Gaussian noise channel, in the achievable region in~\eqref{eq:noPCPTx}, we choose $Q=\emptyset$, we let $V_1,U_1,T_1,U_2,T_2$ be i.i.d. $\mathcal{N}(0,1)$, and
\begin{align*}
   \Xp &= a_1 U_1 + b_1 V_1 + c_1 T_1: \ && |a_1|^2+|b_1|^2+|c_1|^2= 1,
\\ \Xc &= a_2 U_2       +b_2 T_2: \ && |a_2|^2+|b_2|^2= 1.
\end{align*}
With these choices, the channel outputs are
\begin{align*}
  & \Yf = \sqrt{\Cc}  \left(a_1 U_1 + b_1 V_1          + c_1 T_1  \right)+\Zf,
\\& \Yp = \sqrt{\Sp}      \left( a_1 U_1 + b_1 V_1          + c_1 T_1 \right)+\sqrt{\Ic}{\eac}\left( a_2 U_2 + b_2 T_2 \right) +\Zp,
\\& \Yc = \sqrt{\Ip}{\eap}\left( a_1 U_1 + b_1 V_1          + c_1 T_1 \right)+\sqrt{\Sc}      \left( a_2 U_2 + b_2 T_2 \right) +\Zc,
\end{align*}
and the achievable region in~\eqref{eq:noPCPTx} becomes
\begin{subequations}
\begin{align}
  & \Rp \leq   \log \left( \frac{1+\Sp+\Ic}{1+  \Ic |b_2|^2} \right),
\label{eq:c1C1Gaussian}
\\& \Rp 	\leq     \log \left( \frac{1+\Cc}{1+ \Cc \left( |a_1|^2+|c_1|^2\right)} \right) + \log \left( 1 + \frac{\Sp \left( |a_1|^2 + |c_1|^2\right)}{1+ \Ic |b_2|^2} \right),
\label{eq:c2C1Gaussian}
\\& \Rc \leq    \log \left( 1 + \frac{\Sc}{ 1 + \Ip |c_1|^2} \right),
\label{eq:c3C1Gaussian}
\\& \Rp+\Rc	\leq	 \log \left( \frac{1+\Sp+\Ic}{1+  \Ic |b_2|^2} \right) + \log \left( 1 + \frac{\Sc |b_2|^2}{1 + \Ip |c_1|^2} \right),
\label{eq:c4C1Gaussian}
\\& \Rp+\Rc	\leq	\log \left( 1 + \frac{\Sp |c_1|^2}{1 + \Ic |b_2|^2} \right) +  \log \left( \frac{1+ \Sc + \Ip}{1+ \Ip |c_1|^2}\right),
\label{eq:c5C1Gaussian}
\\& \Rp+\Rc	\leq		\log \left( \frac{1+\Cc}{1+ \Cc \left( |a_1|^2+|c_1|^2\right)} \right) + \log \left( \frac{1 + \Sp \left( |a_1|^2 + |c_1|^2\right) + \Ic}{1 + \Ic |b_2|^2}\right) \nonumber
\\& \qquad + \log \left( 1 + \frac{\Sc |b_2|^2}{1 + \Ip |c_1|^2} \right),
 \label{eq:c6C1Gaussian}
\\& \Rp+\Rc	\leq		\log \left( \frac{1+\Cc}{1+ \Cc \left( |a_1|^2+|c_1|^2\right)} \right) + \log \left( 1  +  \frac{\Sp |c_1|^2}{1 + \Ic |b_2|^2} \right)  + \log \left( 1  +  \frac{\Sc + \Ip |a_1|^2}{1+ \Ip |c_1|^2}\right),
 \label{eq:c7C1Gaussian}
\\& \Rp+\Rc	\leq		\log \left( \frac{1+\Cc}{1+ \Cc \left( |a_1|^2+|c_1|^2\right)} \right)+\log \left( \frac{1+\Sp|c_1|^2+\Ic}{1+ \Ic |b_2|^2} \right) \nonumber
\\& \qquad +\log \left( 1 + \frac{\Ip |a_1|^2 + \Sc |b_2|^2}{1+\Ip |c_1|^2}\right), \label{eq:c8C1Gaussian}
\\& 2\Rp+\Rc  \leq		\log \left( \frac{1+\Cc}{1+ \Cc \left( |a_1|^2+|c_1|^2\right)} \right)+\log \left( 1 + \frac{\Sp |c_1|^2}{1 + \Ic |b_2|^2} \right)  + \log \left( \frac{1+\Sp+\Ic}{1+  \Ic |b_2|^2} \right) \nonumber
\\& \qquad +\log \left( 1 + \frac{\Ip |a_1|^2 + \Sc |b_2|^2}{1+\Ip |c_1|^2}\right),
 \label{eq:c9C1Gaussian}
\\& 2\Rp+\Rc	\leq	2 \cdot \log \left( \frac{1+\Cc}{1+ \Cc \left( |a_1|^2+|c_1|^2\right)} \right)+\log \left( 1 + \frac{\Sp |c_1|^2}{1 + \Ic |b_2|^2} \right)    \nonumber
\\& \qquad + \log \left( \frac{1 + \Sp \left( |a_1|^2 + |c_1|^2\right) + \Ic}{1 + \Ic |b_2|^2}\right) +\log \left( 1 + \frac{\Ip |a_1|^2 + \Sc |b_2|^2}{1+\Ip |c_1|^2}\right),
 \label{eq:c10C1Gaussian}
\\& \Rp+2\Rc \leq		\log \left( \frac{1+\Sp|c_1|^2+\Ic}{1+ \Ic |b_2|^2} \right)  + \log \left( 1 + \frac{\Ip |a_1|^2 + \Sc |b_2|^2}{1+\Ip |c_1|^2}\right)+  \log \left( \frac{1+ \Sc + \Ip}{1+ \Ip |c_1|^2}\right), 
\label{eq:c11C1Gaussian}
\\& \Rp+2\Rc \leq	\log \left( \frac{1+\Cc}{1+ \Cc \left( |a_1|^2+|c_1|^2\right)} \right) +\log \left( \frac{1+\Sp|c_1|^2+\Ic}{1+ \Ic |b_2|^2} \right)+ \log \left( 1 + \frac{\Sc |b_2|^2}{1 + \Ip |c_1|^2} \right) \nonumber
\\ & \quad + \log \left( 1 + \frac{\Sc + \Ip |a_1|^2}{1+ \Ip |c_1|^2}\right).
 \label{eq:c12C1Gaussian}
\end{align}
\label{eq:noPCPTxGaussian}
\end{subequations}

\subsection{FME on the achievable rate region when $U_1=\emptyset$} 
\label{sub:noCNCPTx}
We set $U_1=\emptyset$ in the achievable rate region. 
After FME of the achievable region in~\eqref{eq:DPC unified} with $U_1=\emptyset$, we get
\begin{subequations}
\begin{align}
   R_1 & \leq {\rm eq\eqref{eq:c1}} + {\rm eq\eqref{eq:c13}},
\\ R_1 & \leq {\rm eq\eqref{eq:c3}},
\\ R_2 & \leq {\rm eq\eqref{eq:c16}},
\\ R_1+ R_2 & \leq {\rm eq\eqref{eq:c1}} + {\rm eq\eqref{eq:c7}} + {\rm eq\eqref{eq:c18}},
\\ R_1+ R_2 & \leq {\rm eq\eqref{eq:c3}} + {\rm eq\eqref{eq:c18}},
\\  R_1+ R_2 &  \leq {\rm eq\eqref{eq:c2}} 
+ {\rm eq\eqref{eq:c13}} + {\rm eq\eqref{eq:c14}},
\\  R_1+ R_2 & \leq {\rm eq\eqref{eq:c5}} + {\rm eq\eqref{eq:c14}},
\\ R_1 +2R_2 & \leq   {\rm eq\eqref{eq:c4}} + {\rm eq\eqref{eq:c14}} + {\rm eq\eqref{eq:c18}},
\\ R_1 +2R_2 & \leq {\rm eq\eqref{eq:c2}}+ {\rm eq\eqref{eq:c7}}+ {\rm eq\eqref{eq:c14}} + {\rm eq\eqref{eq:c18}},
\\ R_1 +3R_2 & \leq  {\rm eq\eqref{eq:c9}}+ {\rm eq\eqref{eq:c7}}+ {\rm eq\eqref{eq:c14}}+ 2 \cdot {\rm eq\eqref{eq:c18}},
\end{align}
\label{eq:noCNCPTx}
\end{subequations}
for all distributions that factor as~\eqref{eq:input pdf dpc} and by setting $U_1=\emptyset$ in all the mutual information terms.

\begin{remark}
After FME, the following rate constraints (with $U_1=\emptyset$) also appear 
\begin{align*}
R_2 & \leq  {\rm eq\eqref{eq:c11}} + {\rm eq\eqref{eq:c18}},
\\
R_2 & \leq {\rm eq\eqref{eq:c9}} + {\rm eq\eqref{eq:c18}}.
\end{align*}
By following similar steps as in \cite[Lemma 2]{ChongMotaniIC} and \cite[Appendix A]{YANG-TUNINETTI}, it is possible to show that these constraints are redundant. In other words, if these constraints are active a larger rate region is attained by not sending any common message, i.e., by setting $U_2=\emptyset$. 
\end{remark}

We identify Node1 with the PTx (i.e., $\Xp=X_1$), Node2 with the CTx  (i.e., $\Xc=X_2, \Yf=Y_2$), Node3 with the PRx  (i.e., $\Yp=Y_3$) and Node4 with the CRx (i.e., $\Yc=Y_4$). 
For the Gaussian noise channel, in the achievable region in~\eqref{eq:noCNCPTx}, we choose $Q=\emptyset$, we let $S_1,V_1,T_1,Z_1,U_2^{\prime},T_2^{\prime}$ be i.i.d. $\mathcal{N}(0,1)$, and
\begin{align*}
  &\Xp = |a_1|\eac S_1 + b_1 V_1 + c_1 Z_1 + d_1 T_1         \ : |a_1|^2+|b_1|^2+|c_1|^2+|d_1|^2= 1,
\\&\Xc = |a_2|     S_1 + b_2 U_2^{\prime} + c_2 T_2^{\prime} \ : |a_2|^2+|b_2|^2+|c_2|^2= 1,
\\& U_2 = U_2^{\prime}+\lambda_U S_1 \ : \lambda_U = \frac{\Sc|b_2|^2}{\Sc|b_2|^2+\Sc|c_2|^2+1+\Ip (|c_1|^2+|d_1|^2)} \ \frac{\sqrt{\Ip} \eap \eac|a_1|+\sqrt{\Sc}|a_2|}{\sqrt{\Sc} b_2},
\\&T_2 = T_2^{\prime}+\lambda_T S_1 \ : \lambda_T = \frac{\Sc|c_2|^2}{\Sc|c_2|^2+1+\Ip(|c_1|^2+|d_1|^2)} \ \frac{\sqrt{\Ip} \eap \eac |a_1|+\sqrt{\Sc}|a_2| - \sqrt{\Sc} b_2 \lambda_U}{\sqrt{\Sc} c_2},
\end{align*}
where the choice of $\lambda_U$ is so as to ``pre-cancel'' $S_1$ from $\Yc$ in decoding $U_2$,
i.e., so as to have $I(\Yc;U_2|V_1,Q)-I(S_1;U_2|Q)=I(\Yc;U_2|V_1,Q,S_1)$ and the choice of $\lambda_T$ is so as to ``pre-cancel'' $S_1$ from $\Yc$ in decoding $T_2$,
i.e., so as to have $I(\Yc;T_2|V_1,Q,U_2)-I(S_1;T_2|Q,U_2)=I(\Yc;T_2|V_1,Q,U_2,S_1)$.
With these choices, the channel outputs are
\begin{align*}
  & \Yf = \sqrt{\Cc}  \left(|a_1|\eac S_1 + b_1 V_1          + c_1 Z_1 + d_1 T_1 \right)+\Zf,
\\& \Yp = (\sqrt{\Sp} |a_1| +\sqrt{\Ic}|a_2| ) \eac S_1
  +\sqrt{\Sp}      \left( b_1 V_1          + c_1 Z_1 + d_1 T_1\right)+\sqrt{\Ic}{\eac}\left( b_2 U_2^{\prime} + c_2 T_2^{\prime} \right) +\Zp,
\\& \Yc = (\sqrt{\Ip}\eap \eac |a_1|+\sqrt{\Sc}|a_2|)S_1
  +\sqrt{\Ip}{\eap}\left( b_1 V_1          + c_1 Z_1 + d_1 T_1 \right)+\sqrt{\Sc}      \left( b_2 U_2^{\prime} + c_2 T_2^{\prime} \right) +\Zc,
\end{align*}
and the achievable region in~\eqref{eq:noCNCPTx} becomes
\begin{subequations}
\begin{align}
 &\Rp      \leq  \log\left( 1+\frac{\Cc(|b_1|^2+|c_1|^2)}{1+ \Cc |d_1|^2} \right)+ \log\left( 1+ \frac{ \Sp |d_1|^2 }{1+ \Ic |c_2|^2}\right),
\label{eq:c1C2Gaussian}
\\ & \Rp   \leq
\log\left( \frac{1+\Sp+\Ic+2\sqrt{\Sp \Ic |a_1|^2|a_2|^2}}{1+\Ic|c_2|^2}  \right),       
\label{eq:c2C2Gaussian}
\\ & \Rc \leq \log\left( 1+\frac{\Sc(|b_2|^2+|c_2|^2)}{1+ \Ip(|c_1|^2+|d_1|^2)} \right), 
\label{eq:c3C2Gaussian}
\\ & \Rp+ \Rc  \leq
  \log\left( 1+\frac{\Cc(|b_1|^2+|c_1|^2)}{1+ \Cc |d_1|^2} \right)
+ \log \left( 1+\frac{ \Sp |d_1|^2 +  \Ic |b_2|^2}{1+ \Ic |c_2|^2}\right) \nonumber
\\& \qquad + \log\left( 1+\frac{\Sc|c_2|^2}{1+ \Ip(|c_1|^2 +|d_1|^2 )}  \right),
\label{eq:c4C2Gaussian}
\\ & \Rp+ \Rc  \leq   \log\left( \frac{1+\Sp+\Ic+2\sqrt{\Sp \Ic |a_1|^2|a_2|^2}}{1+\Ic|c_2|^2}  \right)
+ \log\left( 1+\frac{\Sc|c_2|^2}{1+ \Ip(|c_1|^2 +|d_1|^2 )}  \right),
\label{eq:c5C2Gaussian}
\\ & \Rp+ \Rc \leq
\log\left( 1+\frac{ \Cc |c_1|^2}{1+ \Cc |d_1|^2} \right) 
+ \log \left(1+\frac{ \Sp |d_1|^2}{1+ \Ic |c_2|^2} \right) \nonumber
\\& \qquad  + \log\left( 1+\frac{\Sc(|b_2|^2+|c_2|^2)}{1+ \Ip(|c_1|^2 +|d_1|^2 )} \right)
+ k_1,
\label{eq:c6C2Gaussian}
\\ & \Rp+ \Rc \leq 
\log \left( 1+\frac{ \Sp (|c_1|^2 +|d_1|^2 )}{1+ \Ic |c_2|^2}\right) + \log\left( 1+\frac{\Sc(|b_2|^2+|c_2|^2)}{1+ \Ip(|c_1|^2 +|d_1|^2 )} \right)
+k_1+k_2,
\label{eq:c7C2Gaussian}
\\ & \Rp +2\Rc  \leq
\log \left( 1+\frac{ \Sp (|c_1|^2 +|d_1|^2 ) +\left(\sqrt{\Sp} |a_1| +\sqrt{\Ic}|a_2|\right)^2+  \Ic |b_2|^2}{1+ \Ic |c_2|^2}\right)
\nonumber
\\& \qquad +\log\left( 1+\frac{\Sc (|b_2|^2+|c_2|^2)}{1+ \Ip(|c_1|^2 +|d_1|^2 )} \right)
+ \log\left( 1+\frac{\Sc|c_2|^2}{1+ \Ip(|c_1|^2 +|d_1|^2 )}  \right)
+k_1,
\label{eq:c8C2Gaussian}
\\ & \Rp +2\Rc \leq 
 \log\left( 1+\frac{\Cc |c_1|^2}{1+ \Cc |d_1|^2} \right) 
+ \log \left( 1+\frac{\Sp  |d_1|^2 + \Ic |b_2|^2 }{1+\Ic|c_2|^2 }\right) \nonumber
\\& \qquad +\log\left( 1+\frac{\Sc (|b_2|^2+|c_2|^2)}{1+ \Ip(|c_1|^2 +|d_1|^2 )} \right)
+ \log\left( 1+\frac{\Sc|c_2|^2}{1+ \Ip(|c_1|^2 +|d_1|^2 )}  \right)
+k_1,
\label{eq:c9C2Gaussian}
\\ & \Rp +3\Rc \leq 
  \log \left( 1+\frac{ \Sp |c_1|^2 +\left(\sqrt{\Sp} |a_1| +\sqrt{\Ic}|a_2|\right)^2+  \Ic|b_2|^2}{1+ \Ic |c_2|^2}\right)+\log \left( 1+\frac{ \Sp |d_1|^2 +  \Ic |b_2|^2}{1+ \Ic |c_2|^2}\right) \nonumber
\\& \qquad +\log\left( 1+\frac{\Sc (|b_2|^2+|c_2|^2)}{1+ \Ip(|c_1|^2 +|d_1|^2 )} \right)
+ 2\log\left( 1+\frac{\Sc|c_2|^2}{1+ \Ip(|c_1|^2 +|d_1|^2 )}  \right)
+k_1,
\label{eq:c10C2Gaussian}
\end{align}
\label{eq:noCNCPTxGaussian}
\end{subequations}
where we defined $k_1:= I \left( \Yc;V_1\right)$ and $k_2:= I \left(\Yp;S_1|V_1,U_2 \right)$ without evaluating them for the Gaussian noise case. Actually, further lower bounding $k_1$ and $k_2$ by zero suffices to prove a constant gap for the yellow and red regions in Fig. \ref{fig:DoF}.

\bibliographystyle{IEEEtran}
\bibliography{ITNewBib}

\end{document}